\documentclass[aps,prb,showpacs,superscriptaffiliation,twocolumn,amsmath,amssymb]{revtex4-1}
\UseRawInputEncoding
\usepackage{graphicx}
\usepackage{dcolumn}
\usepackage{bm}
\usepackage{algorithm}
\usepackage{algpseudocode}
\usepackage{csquotes}
\usepackage{dcolumn}
\usepackage{multirow}
\usepackage[unicode]{hyperref}
\usepackage{longtable}
\usepackage{array}
\usepackage{subfigure}
\usepackage{siunitx}
\usepackage{amsmath}
\usepackage{color}
\usepackage{subfigure}
\usepackage{ifthen}
\usepackage{threeparttable}
\usepackage{float}

\newcommand{\bfr}{ {\bf r}} 
\newcommand{\bfrp}{ {\bf r'}} 
\newcommand{\bfR}{ {\bf R}} 

\newlength\replength

\begin{document}

\title{Localized resolution of identity approach to the analytical gradients of random-phase approximation ground-state energy: algorithm and benchmarks
}

\author{Muhammad N. Tahir}
\affiliation{Guangdong Provincial Key Laboratory of Thermal Management Engineering and Materials, Tsinghua Shenzhen International Graduate School, Tsinghua University, Shenzhen, 518055, China}

\author{Tong Zhu}
       \affiliation{Department of Chemistry, Duke University, Durham, North Carolina 27708, USA}
       
\author{Honghui Shang}
       \affiliation{State Key Laboratory of Computer Architecture, Institute of Computing Technology, Chinese Academy of Sciences, Beijing, 100190, China}

\author{Jia Li}
\email{li.jia@sz.tsinghua.edu.cn}
\affiliation{Guangdong Provincial Key Laboratory of Thermal Management Engineering and Materials, Tsinghua Shenzhen International Graduate School, Tsinghua University, Shenzhen, 518055, China}

\author{Volker Blum}
 \email{volker.blum@duke.edu}
     \affiliation{Department of Chemistry, Duke University, Durham, North Carolina 27708, USA}
     \affiliation{Department of Mechanical Engineering and Materials Science, Duke University, Durham, North Carolina 27708, USA}

\author{Xinguo Ren}
\email{renxg@iphy.ac.cn}
\affiliation{Beijing National Laboratory for Condensed Matter Physics, Institute of Physics, Chinese Academy of Sciences, Beijing 100190, China}

\begin{abstract}

We develop and implement a formalism which enables calculating the analytical gradients of particle-hole random-phase 
approximation (RPA) ground-state energy 
with respect to the atomic positions within the atomic orbital basis set framework. Our approach is based 
on a localized resolution of identity (LRI) approximation
for evaluating the two-electron Coulomb integrals and their derivatives, and the 
density functional perturbation theory 
for computing the first-order derivatives of the Kohn-Sham (KS) orbitals and orbital energies. 
Our implementation allows one to relax molecular structures at the RPA level using both Gaussian-type
orbitals (GTOs) and numerical atomic orbitals (NAOs). Benchmark calculations show that our approach delivers 
high numerical precision compared to previous implementations. A careful assessment of 
the quality of RPA geometries for small
molecules reveals that post-KS RPA systematically overestimates the bond lengths. 
We furthermore optimized the geometries of the four low-lying
water hexamers -- cage, prism, cyclic and book isomers, and determined the energy hierarchy of
these four isomers using RPA. The obtained RPA energy ordering is in good agreement with that yielded by the coupled cluster method with 
single, double and perturbative triple excitations, despite that the dissociation energies themselves 
are appreciably underestimated. The underestimation of the dissociation energies by RPA is well corrected by 
the renormalized single excitation correction.


\end{abstract}

\maketitle

\section{Introduction}
The derivatives of the ground-state total energy with respect to atomic displacements are indispensable in the
computational determination of structural, vibrational, and dynamical properties of molecules and materials. 
Within first-principles approaches, these derivatives are most often obtained by first formally differentiating the energy with respect to the atomic coordinates, and the resulting expression is then evaluated numerically. Such an approach is known
as ``the force method" \cite{pulay:1969} and the derivatives computed this way 
are often called ``analytical gradients". 
For self-consistent variational electronic-structure methods, such as density-functional theory (DFT) in its conventional local, semilocal, or hybrid-functional approximations, the computation of first-order
analytical gradients is particularly simple, thanks to the Hellman-Feynman theorem \cite{Feynman:1939}. This is also one of the key practical advantages underlying the wide-spread use of density functional approximations in computational chemistry and materials science. For non-variational wave-function-based methods such as M{\o}ller-Plesset perturbation theory \cite{Moller/Plesset:1934}, the calculation of analytical gradients is more involved. Nevertheless,
quantum chemists have developed computational formalisms and techniques allowing for routine
evaluations of first- and second-order derivatives of electron correlation energy based 
on non-variational wavefunctions \cite{pulay:1969, doi:10.1002/qua.560270512,doi:10.1002/qua.560160825,doi:10.1002/qua.560270512,doi:10.1002/qua.560160825,doi:10.1063/1.447489,doi:10.1002/qua.560260825,doi:10.1063/1.465552, doi:10.1063/1.468022}.

During the last two decades, the particle-hole random phase approximation (phRPA) \cite{Bohm/Pines:1953,Gell-Mann/Brueckner:1957} within the
framework of the adiabatic connection fluctuation-dissipation theorem \cite{Langreth/Perdew:1977} has been revived as a
promising approach to describe non-local electron correlations in real materials \cite{Hesselmann/Goerling:2011,Eshuis/Bates/Furche:2012,Ren/etal:2012b}. When performed as a post-DFT method that builds on
semi-local or hybrid functional reference states, phRPA is capable of delivering unprecedented accuracy for a wide range of
materials science problems. Recently, an alternative formulation of RPA, namely, 
the particle-particle RPA (ppRPA), corresponding to the ladder channels in the diagrammatic representation, has also received
attentions in quantum chemistry \cite{Aggelen/etal:2013,Scuseria/Henderson/Bulik:2013,Yang/Aggelen/Steinmann/etal:2013,Peng/etal:2014,Tahir/Ren:2019}, but has so far not found as much use as the phRPA.
Below we shall restrict our discussions to the phRPA (simply referred to as RPA in the following) on which 
most present-day application studies are based. Despite its success, the RPA calculations are mostly done
on fixed input geometries, and as such the full potential of the RPA method has not yet been exploited. 

To extend the territory of the applicability of the RPA method, one obvious next step is to calculate the gradients
of the RPA total energy with respect to the atomic positions, and relax the geometries according to the RPA forces.
Developments along these lines have been pioneered by Rekkedal \textit{et al.} \cite{Rekkedal/etal:2013}, 
Burow \textit{et al.}\cite{doi:10.1021/ct4008553}, and Ramberger \textit{et al.}\cite{Ramberger/etal:2017},
whereby the problem was tackled by different formalisms. Specifically, the first reported RPA force implementation of
Rekkedal \textit{et al.} \cite{Rekkedal/etal:2013} was based on the Lagrangian technique \cite{Helgaker/Jorgensen:1992} 
and the equivalence of RPA to ring coupled cluster doubles theory \cite{Scuseria/Henderson/Sorensen:2008}, although
the choice of the Hartree-Fock reference state and the $\mathcal{O}$($N^{6}$) scaling render their 
implementation practically less appealing. By invoking the resolution of identity (RI) approximation and an alternative
formulation of the Lagrangian approach, Burow \textit{et al.}\cite{doi:10.1021/ct4008553} were able to achieve
a canonical $\mathcal{O}$($N^{4}$) scaling of RPA force calculations. Furthermore, using the PBE exchange-correlation functional as the reference state
for RPA calculations, Burow \textit{et al.}\cite{doi:10.1021/ct4008553} found that RPA outperforms the second-order M{\o}ller-Plesset perturbation theory (MP2) for determining the molecular geometries, in particular for transition
metal compounds. Gaussian-type orbitals (GTOs) are used as basis functions in these two implementations. 
In contrast, Ramberger 
\textit{et al.}\cite{Ramberger/etal:2017} developed an elegant formalism which connects the RPA force calculation with 
the $GW$ self-energy \cite{Hedin:1965}, whereby the $O(N^3)$-scaling real-space/imaginary-time $GW$ algorithm
\cite{Liu/etal:2016} can be exploited. This formalism was implemented in the projector augmented wave framework and
allows one to handle periodic systems \cite{Ramberger/etal:2017}. Finally, by employing a similar Green-function based formalism
as Ref.~\cite{Ramberger/etal:2017} and a Cholesky decomposed-density technique, Beuerle and Ochsenfeld \cite{Beuerle/Ochsenfeld:2018}
developed an asymptotically quadratic-scaling algorithm for evaluating the analytical gradients of RPA using atomic orbitals (AOs).


In this work, we present an alternative formalism and implementation of the RPA force gradients, which can be used to relax 
molecular geometries and works both for numerical atom-centered orbitals (NAOs) and GTOs. 
Our implementation is based on
the localized resolution-of-identity (LRI) approximation \cite{Ihrig/etal:2015,Levchenko/etal:2015,Lin/Ren/He:2020} 
(also known as the pair atomic resolution-of-identity (PARI) \cite{Merlot/etal:2013,Wirz/etal:2017}) to represent the pair products of atomic orbitals (AOs) and on
density functional perturbation theory (DFPT) \cite{Baroni/etal:2001,Shang/etal:2017} to compute the derivative 
of the molecular orbitals (MO) with respect to
the nuclear coordinates. Our implementation is carried out within the FHI-aims code package \cite{Blum/etal:2009,Havu/etal:2009,Ren/etal:2012}, and benchmark calculations show that this implementation 
is numerically highly accurate. 
The dependence of the obtained RPA forces on the size of the AO basis sets is examined. We then relax the 
geometries of the different isomers of the water hexamers according to the RPA forces.
The energy ordering of the prism, book, cage, and cyclic isomers of the water hexamers are in excellent agreement
with the high-level coupled cluster method with single, double and perturbative triple excitations (CCSD(T)) \cite{Olson/etal:2007,Biswajit/etal:2008}.

\section{\label{sec:theory}Theoretical Formulation}

In this section, we will first present the key equations behind RPA total energy calculations, based on
the LRI approximation \cite{Ihrig/etal:2015,Levchenko/etal:2015,Lin/Ren/He:2020}, as implemented in FHI-aims. 
After that, we will present the formalism of the gradients of the RPA total energy with respect to the 
nuclear coordinates, highlighting the new terms that need to
be evaluated in the present implementation.

\subsection{RPA total energy within the RI approximation}
In the usual practice of RPA calculations, one first performs a conventional density functional approximation (DFA)
calculation and then evaluates the exact-exchange (EX) energy and RPA correlation energy using single-particle orbitals and orbital energies generated from the preceding DFA calculation. The RPA ground-state total energy is then given by
\begin{equation}
 E^{\text{RPA}}=E^{\text{DFA}}-E^{\text{DFA}}_{xc}+E^{\text{EX}}_{x}+E^{\text{RPA}}_{c}\, ,
 \label{RPA_total_energy}
\end{equation}
where $E^{\text{DFA}}$ is the DFA total energy at the level of, e.g., local-density
or generalized gradient approximations (LDA/GGAs), and $E^\text{DFA}_{xc}$ is the corresponding 
exchange-correlation (xc) energy. Furthermore, $E^{\text{EX}}_{x}$ and $E^{\text{RPA}}_{c}$ are, respectively, the exact-exchange (EX) energy 
and RPA correlation energy evaluated with the DFA orbitals and orbital energies. Specifically,
\begin{equation}
      E^{\text{EX}}_{x}=-\dfrac{1}{2}\sum_{m,n}^{occ} (mn|nm)\, ,
      \label{eq:E_EX}
\end{equation}
where $(mn|nm)$ is the two-electron Coulomb integral in Mulliken's notation,
\begin{equation}
    (mn|nm)=\int \frac{\psi_m(\bfr)\psi_n(\bfr)\psi_n(\bfrp)\psi_m(\bfrp)}{|\bfr-\bfrp|} d\bfr d\bfrp \, .
    \label{eq:ERI}
\end{equation}
Furthermore, the RPA correlation energy can be calculated as \cite{Dobson:1994,Ren/etal:2012b}
\begin{equation}
    E_c^\text{RPA} = \frac{1}{2\pi} \int_0^\infty d\omega \text{Tr}\left[\text{ln}\left(\textbf{1}-\boldsymbol{\chi}^0 (i\omega) \boldsymbol{v}\right) + \boldsymbol{\chi}^0(i\omega) \boldsymbol{v} \right]\,,
\end{equation}
where $\boldsymbol{\chi}^0$ and $\boldsymbol{v}$ should be understood as the matrix form of the noninteracting
density response function and the bare Coulomb interaction, represented in terms of  a set of suitable basis functions. In real space,
$\boldsymbol{\chi}^0$ is given by,
\begin{equation}
\boldsymbol{\chi}^0(\textbf{r},\textbf{r}^{\prime},i\omega) = -\sum_{m,n} \frac{(f_m-f_n)\psi_m(\bfr)\psi_n(\bfr)\psi_n(\bfrp)\psi_m(\bfrp)}{\epsilon_m - \epsilon_n - i\omega}\, .
\label{eq:chi0_realspace}
\end{equation}
In Eqs.~\ref{eq:ERI} and \ref{eq:chi0_realspace}, $\psi_n$, $\epsilon_n$, and $f_n$ are the KS orbitals, their energies and occupation numbers, respectively. For simplicity, here closed-shell systems and real orbitals are assumed. Extension
to spin collinear systems and complex orbitals is straightforward.

Within FHI-aims, the KS MOs are expanded in terms of atom-centered basis functions $\{\varphi_i(\bfr)\}$,
\begin{equation}
    \psi_n(\bfr) = \sum_i c_{i,n} \varphi_i(\bfr-\bfR_I)
\end{equation}
with $c_{i,n}$ being the KS eigenvectors and $\bfR_I$ the position of the atom $I$, to which the basis function $i$ belongs.
Furthermore, the computation of the EX and RPA correlation energies is based on the RI approximation. Within this approximation, 
a set of 
auxiliary basis functions (ABFs) $\{P_\mu(\bfr)\}$ are employed to expand the products of two NAOs, 
\begin{equation}
    \varphi_i(\bfr-\bfR_I)\varphi_j(\bfr-\bfR_J)=\sum_{\mu}C_{ij}^\mu P_\mu(\bfr-\bfR_M)
    \label{eq:RI_AO_expansion}
\end{equation}
where $C_{ij}^\mu$ are the expansion coefficients, and $\bfR_M$ is the position of an atom $M$ on which the ABF $\mu$ is centered. 
In the global RI approximation \cite{Dunlap/Connolly/Sabin:1979,Whitten:1973,Feyereisen/Fitzgerald/Komornicki:1993,Ren/etal:2012}, 
the atom $M$ could be a third atom other than the atoms $I$ and $J$;
however, in the LRI approximation \cite{Ihrig/etal:2015,Levchenko/etal:2015,Lin/Ren/He:2020} adopted in the present work, 
$M$ has to be either $I$ or $J$.
In the RI formalism, another key quantity is the Coulomb matrix, which is the representaton of the Coulomb operator in terms of ABFs,
\begin{equation}
    V_{\mu\nu} = \int d\bfr d\bfrp \frac{P_\mu(\bfr-\bfR_M)P_\nu(\bfrp-\bfR_N)}{|\bfr-\bfrp|}\, .
\end{equation}
Introducing 
\begin{equation}
    Q_{ij}^\mu =\sum_\nu C_{i,j}^\nu \left[V^{\frac{1}{2}}\right]_{\nu,\mu}\,
\end{equation}
and
\begin{equation}
    O_{mn}^\mu = \sum_{ij}c_{i,m}Q_{ij}^\mu c_{j,n}\, ,
    \label{eq:O_integrals}
\end{equation}
it is straightforward to show that
\begin{equation}
E_{x}^{\text{EX}}=-\dfrac{1}{2} \sum_{nm}^{occ} \sum_{\mu}O^{\mu}_{mn} O^{\mu}_{nm} \, .
\label{eq:EX_RI}
\end{equation} 
Further denoting
\begin{equation}
   \mathbf{\Pi}_{\mu\nu} (i\omega)=\sum_{m,n}\frac{(f_m-f_n)O_{mn}^\mu O_{mn}^\nu}{\epsilon_m-\epsilon_n-i\omega}\, ,
   \label{eq:Pi_matrix}
\end{equation}
the RPA correlation energy is then given by
\begin{equation} 
E_c^\text{RPA} = \frac{1}{2\pi} \int_0^\infty d\omega \text{Tr}\left[\text{ln}\left(\textbf{1}-\mathbf{\Pi}(i\omega)\right) + \mathbf{\Pi}(i\omega) \right]\, .
\label{eq:EcRPA_RI}
\end{equation}
More detailed derivations of Eqs.~(\ref{eq:EX_RI}) and (\ref{eq:EcRPA_RI}), on which our RPA force implementation is based, are given in
Refs.~\cite{Ren/etal:2012,Ren/etal:2012b}.

\subsection{RPA gradients within the RI framework}
The RPA force is given by the gradients of the RPA total energy with respect to the atomic displacement
 \begin{equation} \label{eq:RPA_force_total}
 \begin{split}
 \mathbf{F}_{A}^{\text{RPA}}&=-\dfrac{dE^\text{RPA}}{d\mathbf{R}_{A}}\\&
 =-\dfrac{dE^{\text{DFA}}}{d\mathbf{R}_{A}}+\dfrac{dE^{\text{DFA}}_{xc}}{d\mathbf{R}_{A}}-\dfrac{dE^{\text{EX}}_{x}}{d\mathbf{R}_{A}}-\dfrac{dE^{\text{RPA}}_{c}}{d\mathbf{R}_{A}}\, 
 \end{split}
 \end{equation}
 where $\bfR_A$ denotes the nuclear position of a given atom $A$.
In Eq.~\ref{eq:RPA_force_total}, the first term -- the force at the level of  conventional DFAs has long been available in FHI-aims, and only the
remaining three terms are those that need to evaluated in our RPA force implementation. In the following, we will first discuss the EX and RPA correlation parts of
the force, and then turn to the XC part of the conventional DFA force (the second term in Eq.~\ref{eq:RPA_force_total}), to be subtracted from the total DFA force. 
Based on Eqs~(\ref{eq:EX_RI}) and 
\ref{eq:EcRPA_RI}, it is straightforward to show that the force contribution from the EX energy is given by
\begin{equation} 
\mathbf{F}_{x,A}^{\text{EX}}=-\dfrac{dE^{\text{EX}}_{x}}{d\mathbf{R}_{A}} = \sum_{nm} \sum_{\mu}O^{\mu}_{mn} \frac{d O^{\mu}_{nm}}{d\mathbf{R}_{A}}\, , 
\label{eq:EX_force}
\end{equation}
and the force contribution from the RPA correlation energy is
\begin{align}
	F^{\text{RPA}}_{c,A}=-\frac{dE^{\text{RPA}}_{c}}{d\mathbf{R}_{A}}=\dfrac{1}{2\pi}\int \text{Tr}\left[ \mathbf{W}_{c}(i\omega)
	\dfrac{d \boldsymbol{\Pi}(i\omega)}{d\mathbf{R}_{A}} \right] d\omega
	\label{eq:EcRPA_force}
\end{align}  
where 
\begin{equation}
	\mathbf{W}_{c,\mu \nu}(i\omega)=\left[\frac{\boldsymbol{\Pi}(i\omega)}{\textbf{1}-\boldsymbol{\Pi}(i\omega)}\right]_{\mu\nu} \, .
	\label{eq:W_matrix}
\end{equation}
Obviously,  the key step in the RPA force calculation is to evaluate $\dfrac{d\boldsymbol{\Pi}(i\omega)}{d\mathbf{R}_{A}}$, 
formally given by
\begin{equation}
 \begin{split}
 \dfrac{d \boldsymbol{\Pi}_{\mu \nu}(i\omega)}{d\mathbf{R}_{A}}=&\sum_{m,n} (f_m - f_n) O^{\mu}_{mn} \bigg[ \dfrac{2 }{\epsilon_{m}-\epsilon_{n}-i\omega} 
 \dfrac{dO^{\nu}_{mn}}{d\mathbf{R}_{A}}\\&
 -\left(\frac{1}{\epsilon_m - \epsilon_n - i\omega} \right)^2 O^{\nu}_{mn} \left( \dfrac{d \epsilon_m }{d \mathbf{R}_{A}} - \dfrac{d \epsilon_n }{d \mathbf{R}_{A}}  \right)\bigg]\, ,
 \end{split}
 \label{eq:Pi_derivative}
 \end{equation}
 which follows straightforwardly from Eq.~(\ref{eq:Pi_matrix}). 
In Eq.~(\ref{eq:Pi_derivative}), it is evident that one needs to further evaluate the derivatives of the eigenvalues and the 
$O$-integrals with respect to the nuclear coordinates. The derivatives of the
$O$-integrals -- $\dfrac{dO^{\nu}_{mn}}{d\mathbf{R}_{A}}$ -- which are needed
in both Eqs~(\ref{eq:EX_force}) and (\ref{eq:Pi_derivative}), are calculated as

\begin{align} \label{eq:O_derivative}
\dfrac{dO^{\nu}_{mn}}{d\mathbf{R}_{A}~~}=&\sum_{i,j} \left(c_{i,m}Q^{\nu}_{ij}\frac{d c_{j,n}}{d \mathbf{R}_{A}}+  
\frac{d c_{i,m}}{d\mathbf{R}_{A}} Q^{\nu}_{ij} c_{j,n}  \right) \nonumber \\
&+\sum_{ij}c_{i,m}\dfrac{dQ^{\nu}_{ij}}{d\mathbf{R}_{A}~}c_{j,n}\, .
\end{align}

 
In Eq.~(\ref{eq:O_derivative}), the derivatives of $\dfrac{dQ^{\nu}_{ij}}{d\mathbf{R}_{A}}$ only depends on the AOs and the nuclear positions. 
Within FHI-aims, they are evaluated within the LRI approximation, namely, \\
 
 \begin{equation} \label{eq:M_derivative}
\dfrac{dQ^{\nu}_{ij}}{d\mathbf{R}_{A}}=\sum_{\mu}\dfrac{dC^{\mu}_{i,j}}{d\mathbf{R}_{A}}\bigg[V^{\frac{1}{2}}\bigg]_{\mu,\nu} +\sum_{\mu}C^{\mu}_{i,j} \dfrac{d~~}{d\mathbf{R}_{A}}\bigg[V^{\frac{1}{2}}\bigg]_{\mu,\nu} \, .
\end{equation}
Thus, the calculation of RPA forces boils down to the evaluation of the derivatives of the triple $C$ coefficients and the $V$ matrix,
and those of the KS eigenvalues and eigenvectors. Our algorithms for evaluating $\dfrac{dC^{\mu}_{i,j}}{d\mathbf{R}_{A}}$ and 
$\dfrac{d V^{\frac{1}{2}}}{d\mathbf{R}_{A}}$, and for evaluating $\dfrac{d \epsilon_m }{d \mathbf{R}_{A}}$ and
$\dfrac{d c_{i,m}}{d\bfR_A}$ will be discussed in Sec.~\ref{sec:LRI_derivatives} and \ref{sec:DFPT_derivatives}, respectively.

Finally, we consider the force contribution of the XC part of the DFA, which needs to be subtracted to obtain the correct
 RPA total force. This term is calculated as
 \begin{equation}
 \begin{split}
 &\dfrac{d\text{E}^{\text{DFA}}_{xc}}{d \mathbf{R}_{A}}=\int  \dfrac{\delta \text{E}^{\text{DFA}}_{xc}~}{\delta n(\bfr)} \dfrac{\partial n(\bfr)}{\partial \mathbf{R}_{A}} d\bfr = \int {v}_{xc}(\mathbf{r}) \dfrac{\partial n(\bfr)}{\partial \mathbf{R}_{A}} d\bfr\\
 & = 2 \sum_{m}^{occ} \sum_{ij}  c_{i,m}c_{j,m}\int  \ 
 \dfrac{\partial \varphi_{i}(\mathbf{r}-\bfR_I)}{\partial \mathbf{R}{_A}}{v}_{xc}(\mathbf{r}) \varphi_{j}(\mathbf{r}-\bfR_J) d\bfr \\
 & ~~~~ + 2\sum_{m}^{occ}\sum_{ij} \dfrac{d c_{i,m}}{d\bfR_A}c_{j,m} V^{xc}_{ij} \, .
 \end{split}
 \label{eq:force_DFA_xc}
 \end{equation}
 where
    \begin{equation}
   \dfrac{\partial \varphi_{i}(\mathbf{r}-\bfR_I)}{\partial \mathbf{R}{_A}} =
    \begin{cases} \displaystyle
      -\nabla_\bfr \varphi_{i}(\mathbf{r}) ~~~~ I=A \vspace{2pt}  \\ \displaystyle
       0, ~~~~ I\ne A
    \end{cases} \, ,
    \end{equation}
 and
 \begin{equation}
     V^{xc}_{ij} =  \int d\bfr \varphi_{i}(\mathbf{r}) \hat{v}_{xc}(\mathbf{r}) \varphi_{j}(\mathbf{r})\, .
 \end{equation}
 In Eq.~(\ref{eq:force_DFA_xc}), the derivative 
 $\dfrac{d c_{i,m}}{d\bfR_A}$ is again needed, and as alluded
 to above, this will be discussed in Sec.~\ref{sec:DFPT_derivatives}.

 \subsection{Atom pairs and the derivatives of the expansion coefficients within the LRI scheme. }
 \label{sec:LRI_derivatives}
 For a molecular system having $N_{at}$ atoms, there are $N_{at}(N_{at}+1)/2$ many non-equivalent atomic pairs. Within the
 LRI approximation \cite{Ihrig/etal:2015}, the expansion coefficients $C^{\mu}_{ij}$ defined in Eq.~(\ref{eq:RI_AO_expansion}) is given by
\begin{align} \label{eq:expansion_coeff}
C^{\mu}_{ij}= \begin{cases}
\displaystyle\sum_{\substack{\nu\in\mathcal{P}(\mathrm{IJ})\\ I \geq J}} (ij|\nu) L_{\nu \mu}^{IJ}\, ,  \quad \mu \in \mathcal{P}(IJ)\quad \quad \quad \quad \quad  \\
\displaystyle 0\, , \quad \textnormal{otherwise}
\end{cases} 
\end{align}
where $\mathcal{P}(I J)=\mathcal{P}(I)\,\,\bigcup\,\,\mathcal{P}(J)$ specifies the collection of ABFs centering on 
either the atom $I$ or atom $J$ (without losing generality, assuming $I\le J$). 
Furthermore, 
\begin{equation}
(ij|\nu)=\int\frac{\varphi_{i}(\textbf{r}-\bfR_I)\varphi_{j}(\textbf{r}-\bfR_J)P_{\nu}(\textbf{r}^\prime-\bfR_N) }{|\textbf{r}-\textbf{r}^{\prime}|} d\textbf{r} d\textbf{r}^{\prime} \, ,
\end{equation}
and $L^{IJ}=(V^{IJ})^{-1}$ with $V^{IJ}$ being a subblock of the Coulomb matrix with its ABF indices 
associated with the $I$,$J$ atoms, namely,
\begin{equation}
V^{IJ}=
\begin{cases}
 \begin{bmatrix}
  V_{\mu \mu^{\prime}} & V_{\mu \nu} \\
 V_{\nu \mu} & V_{\nu \nu^{\prime}}
 \end{bmatrix},\,\,
\quad \begin{array}{l}\mu,\mu^{\prime}\in \mathcal{P}(\mathrm{I});~ \nu,\nu^{\prime}\in \mathcal{P}(\mathrm{J}) \\
 \textnormal{and} \quad I\ne J \end{array} \\ \\
 \begin{bmatrix}
 V_{\mu\mu^{\prime}}
 \end{bmatrix}
 ,~~~~~~~~~\mu,\mu^{\prime}\in \mathcal{P}(\mathrm{I})~~ \textnormal{and} \quad  I=J \, .
\end{cases}
 \end{equation}
 In essence, the LRI approximation requires that the ABFs to expand an orbital basis 
 product $\varphi_i(\bfr-\bfR_I)\varphi_j(\bfr-\bfR_J)$ restricted
 to the atoms to which the two orbital basis functions belong. It follows that the three-index expansion coefficients 
 $C_{ij}^\mu$ are rather sparse, and their determination can be done separately for each pair of atoms, requiring integrals that involve
 only two atoms.
 
 Now we check how to compute the derivative of the coefficients $C_{ij}^\mu$ with respect to an atomic position $\bfR_A$ --
 $\dfrac{dC^{\mu}_{i,j}}{d\mathbf{R}_{A}}$. Obviously, this derivative is only non-zero for the atom $A$ being either $I$ or $J$.
 Furthermore, in the special case of $I=J$,  the computation of $C_{ij}^\mu$ involves only a single atom and their values don't depend 
 on the geometry of the system. Thus we only need to deal with the $I> J$ case, left with 
 $N_{at}(N_{at}-1)/2$ non-equivalent atom pairs to consider. For each of these atom pairs, the values of $C_{ij}^\mu$ only depends
 on the relative position of the atom $I$ and $J$. We define a vector $\bfR_{IJ}=\bfR_I-\bfR_J$, and the derivatives of $C_{ij}^\mu$ with
 respect to $\bfR_{IJ}$ is then given by,
\begin{equation}
 \begin{split}
 \dfrac{dC^{\mu}_{ij}}{d\textbf{R}_{IJ}}=& 
 -\sum_{\substack{\nu,\nu^{\prime}\in\mathcal{P}(\mathrm{I J}) }}C_{ij}^{\nu}L_{\nu\nu^{\prime}}^{IJ}\dfrac{dV_{\nu^{\prime}\mu}^{IJ}}{d\textbf{R}_{IJ}} 
 +\sum_{\substack{\nu\in\mathcal{P}(\mathrm{IJ})}}\dfrac{d(ij|\nu)}{d\textbf{R}_{IJ}}L_{\nu\mu}^{IJ}
 \end{split}\, \, 
 \end{equation}
 utilizing the property that
 \begin{equation}
     \dfrac{dL^{IJ}}{d\textbf{R}_{IJ}}= \dfrac{d\left(V^{IJ}\right)^{-1}}{d\textbf{R}_{IJ}} = - (L^{IJ})^2  \dfrac{dV^{IJ}}{d\textbf{R}_{IJ}}
 \end{equation}
 The computation of $\dfrac{dC^{\mu}_{ij}}{d\textbf{R}_{IJ}}$ is thus reduced to determining the
 derivatives of the two-center integrals $(ij|\nu)$ and $V_{\mu\nu}^{IJ}$ with respect to the vector
 connecting the two centers (atoms). Note that the derivatives $\dfrac{d(ij|\nu)}{d\textbf{R}_{IJ}}$ and
 $\dfrac{dV_{\mu\nu}^{IJ}}{d\textbf{R}_{IJ}}$ are only three times large in array size as the integrals $(ij|\nu)$ and $V_{\mu\nu}^{IJ}$
 themselves, and hence can be pre-computed and stored in memory.

Finally, the derivatives of $C^{\mu}_{ij}$ with respect to the nuclear coordinates of an arbitrary atom is given by
\begin{align} \label{eq:C_derivative} \displaystyle
 \dfrac{dC^{\mu}_{ij}}{d\textbf{R}_A}=\begin{cases}
 \dfrac{dC^{\mu}_{ij}}{d\textbf{R}_{IJ}}\, ,  \quad\quad\quad  A=I\\ \\ 
 -\dfrac{dC^{\mu}_{ij}}{d\textbf{R}_{IJ}}\, ,  \quad\quad A=J\\  \\
 0\, , \quad\quad\quad\quad\quad\quad \text{else}
 \end{cases}
\end{align}
In Eq.~(\ref{eq:M_derivative}), in addition to $\dfrac{dC^{\mu}_{ij}}{d\textbf{R}_A}$, one also needs to compute
$\dfrac{d V^{\frac{1}{2}}}{d\mathbf{R}_{A}}$, which is straightforwardly given by
\begin{equation}
   \dfrac{d~~}{d\mathbf{R}_{A}}\bigg[V^{\frac{1}{2}}\bigg]_{\mu,\nu} = \frac{1}{2} \sum_{\mu'} \bigg[V^{-\frac{1}{2}}\bigg]_{\mu\mu'} 
    \dfrac{d~~~~}{d\mathbf{R}_{A}}V_{\mu',\nu}\, .
\end{equation}
Here, $V$ is the global Coulomb matrix, and its derivative with respect to the position of a given atom $A$ is a simply collection of
the derivatives of the subblocks of $V^{IJ}$ for individual pairs of atoms $\langle I,J\rangle$. Similarly to Eq.~(\ref{eq:C_derivative}), 
the derivatives of $V^{IJ}$ are given by
\begin{align} \label{eq:V_IJ_derivative} \displaystyle
 \dfrac{dV^{IJ}_{\mu\nu}}{d\textbf{R}_A}=\begin{cases}
 \dfrac{dV^{IJ}_{\mu\nu}}{d\textbf{R}_{IJ}}\, ,  \quad\quad\quad  A=I \, .\\ \\ 
 -\dfrac{dV^{IJ}_{\mu\nu}}{d\textbf{R}_{IJ}}\, ,  \quad\quad A=J \, .\\  \\
 0\, , \quad\quad\quad\quad\quad\quad \text{else}\, .
 \end{cases}
\end{align}

\subsection{DFPT within the NAO basis set}
 \label{sec:DFPT_derivatives}
In our formulation, the remaining quantities still need to be evaluated are the derivatives of the eigenvalues and eigenvectors with respect
to the nuclear coordinates. As mentioned above, these are obtained via the DFPT approach \cite{Baroni/etal:2001}.
The implementation details of DFPT in FHI-aims have been presented in Ref.~\cite{Shang/etal:2017}, and here we only 
recapitulate the key formulae that are essential for RPA force calculations in our implementation.

To understand how to determine $\dfrac{d \epsilon_m }{d \mathbf{R}_{A}}$ and $\dfrac{d c_{i,m}}{d\bfR_A}$ via
DFPT, one begins with the the KS equations expressed in terms of a finite NAO basis representation,
\begin{equation} \label{eq:KS_eq}
\sum_{j} H_{ij}[\{\bfR\}] c_{j,m}[\{\bfR\}]=\epsilon_{m}[\{\bfR\}]\sum_{j} S_{ij}[\{\bfR\}] c_{j,m}[\{\bfR\}]\, .
\end{equation}
where $H_{ij} =\langle \varphi_{i} |\hat{h}_\text{KS} |\varphi_{j}\rangle$ and $S_{ij}=\langle \varphi_{i} | \varphi_{j} \rangle$  
are the Hamiltonian and overlap matrices, respectively.
In Eq.~\ref{eq:KS_eq}, we have explicitly indicated that
$H_{ij}$ and $S_{ij}$, as well as the eigenvalues $\epsilon_{m}$ and eigenvectors $c_{j,m}$ are all functions of
the atomic positions $\{\bfR\}$ in the system. At a given geometrical configuration $\{\bfR^{(0)}\}$, taking 
the derivatives of Eq.~\ref{eq:KS_eq} with respect to the coordinates of a given
atom $A$ (i.e., $\bfR_A$), one obtains,
\begin{equation} \label{eq:sternheimer_equation}
   \begin{split}
& \sum_{j}\Big( H^{(0)}_{ij}-\epsilon^{(0)}_{m}S^{(0)}_{ij}\Big)c^{(1)}_{j,m}= \\ &
   \epsilon_{m}^{(0)}\sum_{j}S^{(1)}_{ij}c^{(0)}_{j,m}-\sum_{j}\Big( H^{(1)}_{ij}-\epsilon_{m}^{(1)}S^{(0)}_{ij}\Big )c^{(0)}_{j,m}\, .
   \end{split}
\end{equation}  
Here for simplicity, we have denoted $H^{(0)}_{ij}=H_{ij}[\{\bfR^{(0)}\}]$, $S^{(0)}_{ij}=S_{ij}[\{\bfR^{(0)}\}]$, $\epsilon^{(0)}_{m}=\epsilon_{m}[\{\bfR^{(0)}\}]$,
$c^{(0)}_{j,m}=c_{j,m}[\{\bfR^{(0)}\}]$, and $H^{(1)}_{ij}=\dfrac{dH_{ij}[\{\bfR\}]}{d\bfR_A}\Big|_{\{\bfR^{(0)}\}}$, $S^{(1)}_{ij}=\dfrac{dS_{ij}[\{\bfR\}]}{d\bfR_A}\Big|_{\{\bfR^{(0)}\}}$, $\epsilon^{(1)}_{m}=\dfrac{d\epsilon_{m}[\{\bfR\}]}{d\bfR_A}\Big|_{\{\bfR^{(0)}\}}$,
and $c^{(1)}_{j,m}=\dfrac{dc_{j,m}[\{\bfR\}]}{d\bfR_A}\Big|_{\{\bfR^{(0)}\}}$. For a $N$-atom system, Eq.~\ref{eq:sternheimer_equation} represents 3$N$
independent linear equations (also known as Sternheimer equation~\cite{sternheimer:1954}), from which $c^{(1)}_{j,m}$ and $\epsilon^{(1)}_{m}$ can be obtained. The first-order derivatives $H^{(1)}_{ij}$ and $S^{(1)}_{ij}$ that are needed in Eq.~\ref{eq:sternheimer_equation}, can be obtained
via efficient grid-based integrations, and have been available in FHI-aims \cite{Blum/etal:2009,Knuth/etal:2015,Shang/etal:2017}. 
It is also customary to express the first-order derivatives $c^{(1)}_{j,m}$ in terms of the zeroth-order eigenvectors,
\begin{equation}
    c^{(1)}_{j,m} = \sum_n c^{(0)}_{j,n} U_{nm}^{(1)}\, ,
\end{equation}
where $U_{nm}^{(1)}$ are the expansion coefficients. From Eq.~\ref{eq:sternheimer_equation}, it is straightforward to obtain
   \begin{equation} \label{eq:DFPT_U_matrix}
    U_{nm}^{(1)}=
    \begin{cases} \displaystyle
     \sum_{ij} c_{i,n}^{(0)}\dfrac{\epsilon_{n}^{(0)} S^{(1)}_{ij}-H^{(1)}_{ij} }{\epsilon_{n}^{(0)}-\epsilon_{m}^{(0)}}c_{j,m}^{(0)}~, ~~ n\ne m \vspace{7pt}  \\ \displaystyle
     -\dfrac{1}{2}\sum_{ij}c_{i,n}^{(0)}S^{(1)}_{ij}c_{j,n}^{(0)}~, ~~~~~~~~~~ n=m
    \end{cases}
    \end{equation}
   and 
   \begin{equation*}
   \epsilon_{m}^{(1)}=\sum_{ij}c_{i,m}^{(0)}\left[H^{(1)}_{ij}-\epsilon_{m}^{(0)} S^{(1)}_{ij}\right]c_{j,m}^{(0)}\, .
   \end{equation*}

Using Eq.~\ref{eq:DFPT_U_matrix}, the calculation of $\dfrac{dO^{\nu}_{mn}}{d\mathbf{R}_{A}~~}$ in Eq.~\ref{eq:O_derivative}
can be simplified as 
\begin{align} \label{O_derivative}
\dfrac{dO^{\nu}_{mn}}{d\mathbf{R}_{A}~~}=\sum_{l} \big(O^{\nu}_{ml}U^{(1)}_{ln}+ O^{\nu}_{lm}U^{(1)}_{ln} \big)+\sum_{ij}c_{i,m}\dfrac{dM^{\nu}_{ij}}{d\mathbf{R}_{A}~}c_{j,n}\, .
\end{align}

So far we have discussed all the ingredients that are needed for RPA force calculations in our implementation. In Algorithm~\ref{alg:RPA_force_algorithm}, 
we present a flowchart which illustrates the major steps for evaluating the RPA gradients. From the flowchart, we can see that the dominating steps are 
the calculations of the derivatives of the $O$-integrals (cf. Eq.~\ref{eq:O_derivative}) and the $\Pi$-integrals (cf.~\ref{eq:Pi_derivative}), both scaling
as $O(N^5)$ with respect to the system size $N$ (in case of $\dfrac{\Pi}{\bfR_A}$, the $O(N^5)$ scaling estimate assumes 
that the number of frequency points $N_\omega$ does not increase with $N$). Such a high scaling exponent
results from the MO representation of the two dominating steps noted above. The advantage of this algorithm is 
that it has a small prefactor and its implementation 
is relatively straightforward, but its $O(N^5)$ scaling prevents one from going to large systems. Switching from the MO to AO representation, and exploiting the sparsity 
owing to the locality of AOs and the LRI approximation, one can derive a more advanced low-scaling algorithm for RPA force calculations, 
but this has not been implemented and hence will not be discussed here. For the reminder of this paper, the focus will be to demonstrate
the correctness and precision of the general expressions outline above.

%

\begin{figure*}
	\begin{minipage}{0.7\linewidth}
		\begin{algorithm}[H]
	\label{alg:RPA_force_algorithm}
	\caption{ Flowchart for evaluating the derivatives of the RPA correlation energy with respect to nuclear coordinates.
	Here $N_{at}$, $N_b$, $N_{occ}$, $N_{unocc}$, $N_{aux}$, $N_{\omega}$,  are the numbers of atoms, the AO basis functions,
	the occupied states, the unoccupied states, the ABFs, and the frequency points, respectively. The scaling behaviors
	of the major steps are indicated on the right.
	}
	\begin{algorithmic}[1]
		\vspace{4pt}
   \Statex \hspace{7.5cm}	\textbf{Scaling Behavior}
    \vspace{8pt}
	\State Calculate $c_{i,m}$ and $\epsilon_{m}$ from SCF cycle ~~~~~~~~~~~~~~~$O(N_{b}^3)$
\vspace{8pt}
 \State Determine $c_{i,m}^{(1)}$ and $\epsilon_{m}^{(1)}$ from DFPT~(cf. Eq.~\ref{eq:sternheimer_equation}) ~$O(3\times N_{at}N_{b}^3)$
 \vspace{8pt}
 \State Construct $U^{(1)}_{nm}$   ~~~(cf. Eq.~\ref{eq:DFPT_U_matrix})~~~~~~~~~~~~~~~~~~~~~~~~~~ $O(3\times N_{at}N_{b}(N_{occ}+N_{unocc})^2)$
  \vspace{8pt}
		\State Construct $O^{\nu}_{mn}$ ~~ (cf. Eq.~\ref{eq:O_integrals}) ~~~~~~~~~~~~~~~~~~~~~~~~~~$O(N_{aux}N_{b}^2(N_{occ}+N_{unocc}))$
		\vspace{8pt}
		\State Compute $\dfrac{dO^{\nu}_{mn}}{d\bf{R}_A~~}$ ~~ (cf. Eq.~\ref{eq:O_derivative})  ~~~~~~~~~~~~~~~~~~~~~~~~~~~~~~~~~~~~~~~~$O(3\times N_{at}N^{2}_{b}(N_{occ}+N_{unocc})N_{aux})$
		\vspace{8pt}
		\State $F^{\text{RPA}}_{c,A}=0$
		\vspace{4pt}
		\For {$\omega \gets 1$ to $N_\omega$}
		  	\vspace{8pt}
		  	\State Compute $W_{c}^{\mu\nu}(i\omega)$ ~~ (cf. Eq.~\ref{eq:W_matrix})  ~~~~~~~~~~~~~~~~~~~~ $O(N^{3}_{aux}N_{\omega})$
		  	\vspace{4pt}
		  	\State Evaluate $\dfrac{d\Pi(i\omega)_{\mu\nu}}{d\bf{R_{A}}~~~~}$   ~~~(cf. Eq.~\ref{eq:Pi_derivative}) ~~~~~~~~~~~~~~~~~~~~~~~$O(3\times N_{at}N^{2}_{aux}N_{occ} N_{unocc}N_{\omega})$
		  	\vspace{4pt}
		  	\State $F^{\text{RPA}}_{c,A} \gets 
		  	      F^{\text{RPA}}_{c,A} + \text{Tr}\left[W_{c}(i\omega)\dfrac{d\Pi(i\omega)}{d\bf{R}_{A}}\right]$(cf. Eq.~\ref{eq:EcRPA_force})~~$O(3\times N_{at}N_{aux}^{3}N_{\omega})$
		  	
		\vspace{4pt}
		\EndFor
		\vspace{4pt}
	    \State $F^{\text{RPA}}_{c,A} \gets F^{\text{RPA}}_{c,A}/2\pi$ 
	 \end{algorithmic}
	\end{algorithm}
	\end{minipage}
\end{figure*}

\section{Computational details and validation of the correctness of the implementation}
Our RPA gradient formalism described above has been implemented in the FHI-aims code package \cite{Blum/etal:2009,Havu/etal:2009,Ren/etal:2012}, 
which allows one
to use both NAOs and GTOs as basis functions. Consequently, our implementation also works for both types of basis functions.
For a given set of AO basis functions, the ABFs are generated on the fly according to the automatic procedure described 
in Refs.~\cite{Ren/etal:2012,Ihrig/etal:2015}, with a threshold of $10^{-2}$ used for the Gram-Schmidt orthonormalization scheme
(the keyword \textit{prodbas\_acc} in FHI-aims). A modified Gauss-Legendre frequency grid with 24 points is used
for the frequency integration in Eq.~\ref{eq:EcRPA_force} \cite{Ren/etal:2012}.

To check the correctness of our implementation, we have performed two sets of validation checks. As a first check, we compare the RPA forces obtained by the 
analytical gradient approach and the finite difference (FD) method for a set of diatomic molecules at fixed bond lengths. In these calculations,
we used the NAOs with FHI-aims's ``tight" settings for the basis size, basis cutoff radius, and numerical integration grid. 
The FD RPA force is calculated as
\begin{equation}
    F_\text{FD}(x)= \dfrac{E^\text{RPA}(x+h)-E^\text{RPA}(x-h)}{2h}
\end{equation}
where $x$ is the bond length of the molecule and $h$ is set to 0.001 {\AA}. The obtained results for a set of diatomic molecules are presented
in Table~\ref{tab:finite_difference_tight_default}, from which one can see that the RPA forces calculated by the analytical gradient technique
agree very well with those obtained by the FD method. In all cases, the differences are around 1  meV/\AA, and the relative deviation is about 1\%. 
This is an acceptable level of discrepancy since we don't expect a perfect agreement, given that the FD difference results also contain higher-order effects. 
For comparison, we also present in Table~\ref{tab:finite_difference_tight_default} the analytical and FD forces 
at the PBE level for the same set of molecules, and a similar level of differences is observed. In particular, the bond lengths of the molecules in
Table~\ref{tab:finite_difference_tight_default} are
chosen somewhat arbitrarily, and consequently the magnitude of the forces spreads over a large range. Thus the level of agreement between analytical 
and FD RPA forces is considered to be rather satisfactory. This benchmark test clearly shows that our RPA gradient implementation is internally consistent.

\begin{table*} [ht]
	\caption{\label{tab:finite_difference_tight_default}The RPA@PBE and PBE forces (labeled as $F$, in eV/\AA ) for a set of diatomic molecules 
	calculated using the analytical gradient technique 
	developed in this work and the finite difference (FD) method (labelled as $F_\text{FD}$). The ``\textit{tight}" NAO basis set is used for
	both the PBE and RPA@PBE calculations. Furthermore, the frozen-core approximation is adopted for RPA correlation energy calculations.
	The ``Dist" column lists the bond lengths of the molecules at which the forces are calculated. The atoms are displaced by $\pm 0.001$ {\AA} 
	around the chosen bond length for the FD calculations. For each molecule, the difference $\Delta F=F-F_\text{FD}$ and absolute percentage 
	deviation (APD) is given by $|\Delta F|/F_\text{FD}*100\%$. The mean absolute percentage deviation (MAPD) is given by the average of APD for
	all molecules.}
	\begin{tabular}{lcrrrrcrrrr} 
		\hline\hline 
		\multicolumn{1}{c}{\multirow{2}{*}{System}} & \multicolumn{1}{c}{\multirow{2}{*}{Dist.(\AA)}} & \multicolumn{4}{c}{RPA@PBE} & & \multicolumn{4}{c}{PBE} \\
		\cline{3-6} \cline{8-11} 
		&  &\multicolumn{1}{c}{\multirow{1}{*}{$F$ (eV/\AA)}} & \multicolumn{1}{c}{\multirow{1}{*}{$F_\text{FD}$ (eV/\AA)}} & \multicolumn{1}{c}{\multirow{1}{*}{$\bigtriangleup F$ (eV/\AA)}}& \multicolumn{1}{c}{\multirow{1}{*}{ APD }}    
		& &\multicolumn{1}{c}{\multirow{1}{*}{$F$ (eV/\AA)}} & \multicolumn{1}{c}{\multirow{1}{*}{$F_\text{FD}$ (eV/\AA)}}&  \multicolumn{1}{c}{\multirow{1}{*}{$\bigtriangleup F$ (eV/\AA)}} & \multicolumn{1}{c}{\multirow{1}{*}{APD}} \\ 
		\hline 
		H$_{2}$  & 0.7496  &  0.180035& 0.178945 & 0.001090 & 0.61 \% &  &0.018787 &0.019250 & $-$0.000463& 2.41 \%\\
		He$_{2}$  & 5.2740 & 0.001362& 0.001430&$-$0.000068& 4.76 \%& &0.000022 & 0.000020 & 0.000002 & 10.00 \%\\
		Li$_{2}$  & 2.6257  &0.106865 & 0.105540  & 0.001325&1.26 \%  & & 0.149533 & 0.147925 & $-$0.001608& 1.09 \%\\
		Be$_{2}$  & 2.5000  & 0.070556 &0.069895& 0.000661& 0.95 \% & & 0.116697 &0.115815 & 0.000882& 0.76 \%\\
		N$_{2}$  & 1.1220  & 1.649008 &1.650010 & $-$0.001002 & 0.06 \% & &2.568019 & 2.567975 & 0.000044 & 0.00 \%\\
		F$_{2}$  & 1.000  & 58.133551 & 58.136530 & $-$0.002979 & 0.01 \% & &57.841913 & 57.841960 & $-$0.000047 & 0.00 \% \\
		Ne$_{2}$  & 2.6462  & 0.076480 & 0.077910 & $-$0.001430 & 1.84 \% & &0.081419 & 0.081220 & 0.000199 & 0.25 \% \\
		Na$_{2}$   &3.0790  & 0.130134 &0.131075 & $-$0.000941 & 0.72 \% & &0.011107&0.011395 & $-$0.000288& 2.53 \% \\
		HF  & 1.0000 &3.618271 &3.616880 & 0.001391 & 0.04 \% & &3.027647 &3.027790 & $-$0.000143& 0.01 \%\\
		BF  & 1.2570  & 0.572137 & 0.574815& $-$0.002678 & 0.47 \% & &0.794657 & 0.796380 & $-$0.001723& 0.22 \%\\
		CO  & 1.0000  &25.881483 & 25.881340 & 0.000143 &0.00 \%  & & 25.472393  & 25.472810 &0.000417 & 0.00 \%\\
		$\text{NO}^{+}$  & 1.0656  & 1.1381667& 1.140335 &$-$0.002168 & 0.19 \% & & 0.626109 & 0.628580 & $-$0.002471 & 0.39 \%\\
		HCl   & 2.0000 & 3.731035 &3.730435 & 0.000600& 0.02 \% & & 3.963126 &3.961520 & 0.001606 & 0.04 \%\\
		Cl$_{2}$  & 1.9878 & 0.651571 &0.652655 & $-$0.001084& 0.17 \% &  & 0.336947 & 0.336595 & 0.000352& 0.11 \%\\ [2.0ex]
		MAE & & & & 0.001254 & & & & &0.0007316  \\
	 MAPD & & & & & 0.79 \% & & & & & 1.27 \%   \\
		\hline\hline
	\end{tabular}
\end{table*}

As a second check of the reliability of our implementation, we compared  for a set of molecules the relaxed RPA geometries obtained in this work 
with those reported by 
Burow \textit{et al.} in Ref.~\onlinecite{doi:10.1021/ct4008553}. In Table.~\ref{tab:comparison_wd_Burow}, the RPA@PBE and PBE bond lengths and angles 
obtained in the present work, together with the literature values, are presented. The same (PBE) reference states and basis set (def2-QZVPPD) are
used in the RPA calculations, and hence the present results and the literature values in Ref.~\onlinecite{doi:10.1021/ct4008553} are directly
comparable. 
Table~ \ref{tab:comparison_wd_Burow} shows that the differences in bond lengths between Ref.~\cite{doi:10.1021/ct4008553} and the present implementation 
within FHI-aims are around 10$^{-2}$ pm, and the difference in bond angles are around 0.01$^\circ$, with the sole exception of the H-N-N angle in the N$_2$H$_2$.
However, in this case, the appreciably big difference of $2.6^\circ$ in the RPA angles is already present at the PBE level, and does not stem from the
inaccuracy in the RPA gradient implementation. We then performed a PBE relaxation with a third code -- NWChem \cite{nwchem} for N$_2$H$_2$, and obtained
a bond angle of 106.4510$^\circ$, in close agreement with the FHI-aims value of 106.4440$^\circ$. This indicates there is probably an error regarding
the bond angle of N$_2$H$_2$ in Ref.~\cite{doi:10.1021/ct4008553}. Excluding this exceptional case, the obtained RPA bond lengths and angles for
this set of molecules agree up to a very high numerical precision within the two implementations. 
This is quite remarkable considering the rather different theoretical formalisms and RI schemes employed behind the two implementations. 

\begin{table*} [ht!]
	\caption{\label{tab:comparison_wd_Burow} Comparison of the calculated equilibrium bond lengths $R_{e}$ (in pm) and bond angles $\varTheta_{e}$ (in deg) based on the RPA@PBE and PBE optimized geometries obtained in this work and reported in Ref.~\onlinecite{doi:10.1021/ct4008553}. All 
	calculations were done with the QZVPPD \cite{doi:10.1063/1.3484283} basis sets. 
	The \textquotedblleft Difference\textquotedblright column presents the differences between the RPA@PBE and PBE bond lengths/angles obtained in 
	this work using FHI-aims and those taken from Ref.~\onlinecite{doi:10.1021/ct4008553}. Frozen-core approximation is not used in the
	present RPA calculations.}
	\begin{tabular}{lcrrrrrr} 
		\hline\hline 
		\multicolumn{1}{c}{\multirow{2}{*}{System}} &
		\multicolumn{1}{c}{\multirow{2}{*}{Length $\&$ angle}}  & \multicolumn{1}{c}{\multirow{2}{*}{\small{RPA@PBE}}} & \multicolumn{1}{c}{\multirow{2}{*}{\small{RPA@PBE}}} & \multicolumn{1}{c}{\multirow{2}{*}{Difference}} & \multicolumn{1}{c}{\multirow{2}{*}{~~~~~PBE}}  & \multicolumn{1}{c}{\multirow{2}{*}{~~~~~PBE}} & \multicolumn{1}{c}{\multirow{2}{*}{Difference}}\\[2.0 ex]
		& & (Ref.~\onlinecite{doi:10.1021/ct4008553}) & (this work) & & 
		(Ref.~\onlinecite{doi:10.1021/ct4008553}) & (this work) & \\
		\hline \\ [-0.5ex]
		\multicolumn{8}{c}{Bond lengths} \\ [3ex]
		H$_{2}$ & $R_{e}$  &74.2891~ &74.3247~ & $-$0.0356~ & 74.9691~ &75.0074~ & $-$0.0383~ \\
		F$_{2}$ & $R_{e}$ &  143.4980~ & 143.4801~ & 0.0179~  & 141.3180~ &141.2986~ & 0.0194~ \\
		N$_{2}$ & $R_{e}$ &  110.4030~ & 110.4024~ & 0.0006~  &110.2030~ &110.2012~ &  0.0018~  \\
		HF & $R_{e}$ & 92.0179~ & 92.0372~ & $-$0.0193~  & 93.0079~ &93.0197~ & $-$0.0118~ \\
		CO & $R_{e}$  &113.5460 ~ & 113.5414~ &0.0046~ & 113.5360~ &113.5304~ & 0.0056~ \\
		CO$_{2}$ & $R_{e}$ & 116.6460~ & 116.6314~ & 0.0146~ & 117.0640 & 117.0445~ &0.0195~  \\
		\multirow{1}{*}{\text{HCN}}   & $R_{e}$ (H-C)  & 106.6580~ & 106.6500~ & 0.0080~ & 107.4880~ & 107.4820~ & 0.0060~  \\ 
		& $R_{e}$ (C-N)  & 115.8460~ &115.8310 ~& 0.0150~ & 115.7460~ &115.7420~ & 0.0040~ \\
		\multirow{1}{*}{\text{HNC}}   & $R_{e}$ (H-N) & 99.5490 & 99.5549~ & $-$0.0059~ & 100.4990~ &100.5130~ & $-$0.0140~ \\ 
		& $R_{e}$ (C-N)  & 117.5150~ & 117.5140~ & 0.0010~ &117.4750~ &117.4700~ & 0.0050~ \\
		
		
		\multirow{1}{*}{\text{H$_{2}$O}}   & $R_{e}$ (H-O)  & 96.1202~ & 96.1314~ & $-$0.0112~ &96.8802~ & 96.8992~ & $-$0.0190~ \\ 
		
		\multirow{1}{*}{\text{HNO}}   & $R_{e}$ (N-O)  & 121.5390~ & 121.5300~ & 0.0090~ & 120.6890~ & 120.6900~ & $-$0.0010~ \\ 
		& $R_{e}$ (N-H)  & 105.3490~ & 105.3710~ & $-$0.0220~ & 107.9690~ & 107.9950~ & $-$0.0260~ \\
		
		\multirow{1}{*}{\text{HOF}}   & $R_{e}$ (H-O) & 97.0219~ & 97.0353~ & $-$0.0134~ & 97.9619~ & 97.9795~ &  $-$0.0176~ \\ 
		& $R_{e}$ (O-F)  &145.5570~ & 145.5580~ & $-$0.0010~ &144.4870~ &144.4830~ & 0.0040~ \\
		
		\multirow{1}{*}{\text{NH$_{3}$}}   & $R_{e}$ (H-N)  & 101.3390~ & 101.4180~  &$-$0.0790~ & 102.0990~ & 102.1220~ & $-$0.0230~ \\ 
		
		
		\multirow{1}{*}{\text{H$_{2}$S}}   & $R_{e}$ (H-S) & 133.6600~ & 133.7000~ &0.0400~ & 135.1100~ & 135.1430~ & $-$0.0330~	\\
		
		\multirow{1}{*}{\text{N$_{2}$H$_{2}$}}   & $R_{e}$ (H-N) & 103.0930~ & 103.1140~ & $-$0.0210~ & 104.3530~ & 104.3750~ &   $-$0.0220~ \\
		& $R_{e}$ (N-N)  & 125.0550~  & 125.0520~ & 0.0030~  &124.5850~ & 124.5900~ & $-$0.0050~ \\
		
		\multirow{1}{*}{\text{CH$_{2}$O}}   & $R_{e}$ (C-O)  & 120.9550~ & 120.9480~ &~ 0.0070~ & 120.7650~ & 120.7690~ & $-$0.0040~ \\
		& $R_{e}$ (C-H) & 110.1920~ & 110.1950~ &$-$0.0030~ & 111.7020~ & 111.7230~ & $-$0.0210~ \\
		MAE & & & & 0.0158~ & & & 0.0143~ \\

		\hline \\[-0.5ex]
		\multicolumn{8}{c}{Bond angles} \\ [3ex]	
		\text{CH$_{2}$O}   &$\varTheta_{e}$ (HCO)  & {121.8100}$^\circ$ & {121.8169}$^\circ$&$-${0.0069}$^\circ$ &{122.0300}$^\circ$ & {122.0243}$^\circ$ & {0.0057}$^\circ$ \\
		H$_{2}$O  & $\varTheta_{e}$  &{103.9300}$^\circ$ & {103.9265}$^\circ$& {0.0035}$^\circ$ &{104.2100}$^\circ$ &{104.2060}$^\circ$ &{0.0040}$^\circ$ \\
		HNO & $\varTheta_{e}$  &{108.2300}$^\circ$ & {108.2314}$^\circ$& {0.0014}$^\circ$ & {108.7600}$^\circ$ &{108.7631}$^\circ$ & $-${0.0031}$^\circ$\\ 
	    HOF & $\varTheta_{e}$  &{97.2300}$^\circ$ & {97.2448}$^\circ$& $-${0.0148}$^\circ$ &{98.0300}$^\circ$ & {98.0151}$^\circ$& {0.0149}$^\circ$ \\
		 NH$_{3}$ & $\varTheta_{e}$  & {106.1600}$^\circ$& {106.1724}$^\circ$ & $-${0.0124}$^\circ$ &{106.2900}$^\circ$ &{106.2737}$^\circ$ & {0.0163}$^\circ$\\
		 H$_{4}$S &$\varTheta_{e}$  & {92.0700}$^\circ$ & {92.0926}$^\circ$&$-${0.0226}$^\circ$ &{91.6900}$^\circ$ & {91.6917}$^\circ$& $-${0.0017}$^\circ$\\
		 N$_{2}$H$_{2}$ &$\varTheta_{e}$   & {103.54}$^\circ$ & {106.0920}$^\circ$&  $-${2.5520}$^\circ$ & {103.7900}$^\circ$& {106.4440}$^\circ$ & $-${2.6540}$^\circ$ \\
		 MAE & & & & 0.3734$^{\circ}$ & & & 0.3857$^{\circ}$ \\
		\hline\hline
		
	\end{tabular}
\end{table*}

In summary, the comparison to the FD results and the RPA geometries reported in the literature validates our RPA analytical gradient implementation.

\section{Results and Discussion}

\subsection{RPA equilibrium geometries for small molecules}
\begin{figure}[ht!]
	\centering
	\begin{minipage}{0.49\textwidth}
		\includegraphics[width=1.\textwidth]{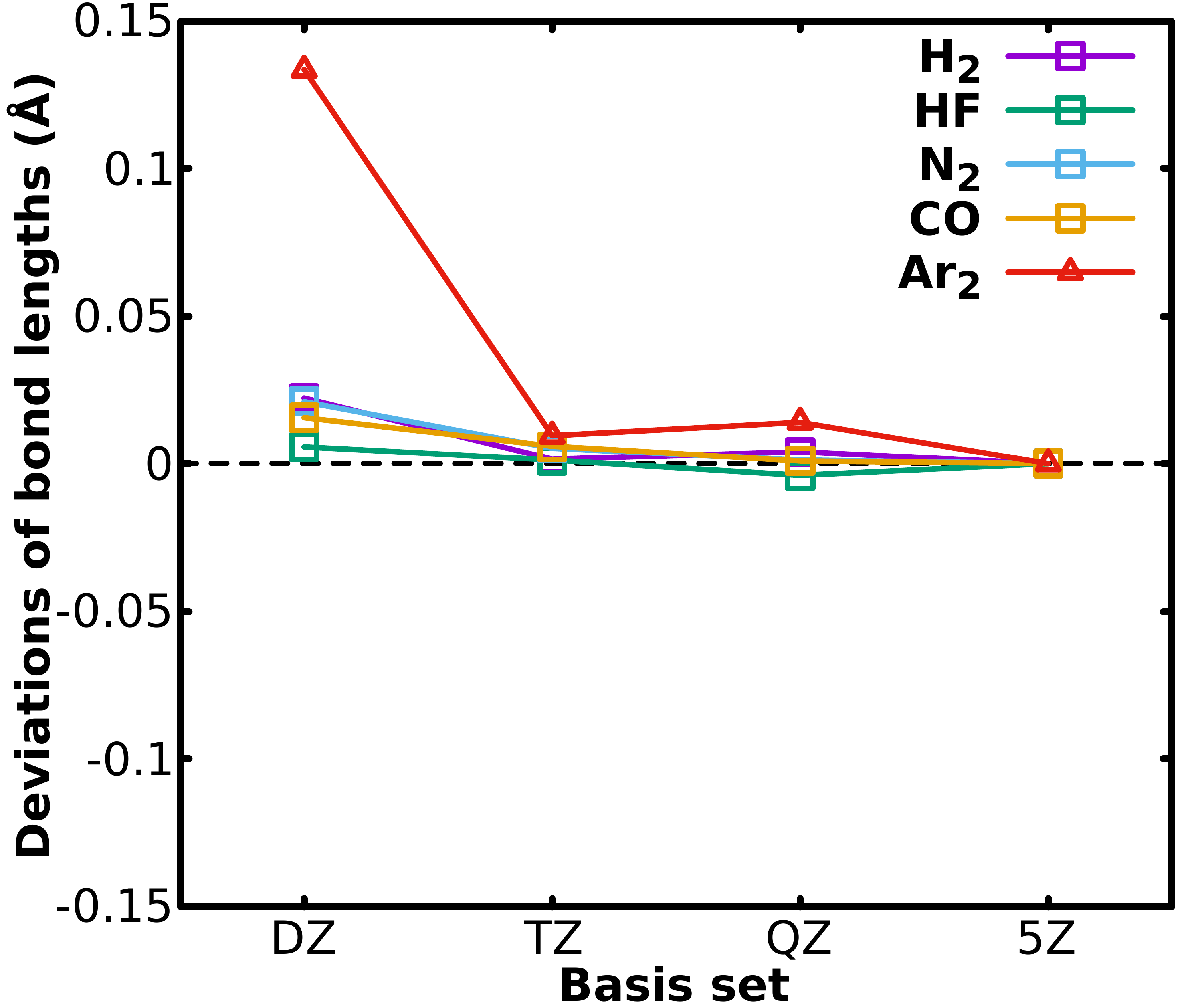} 
	\end{minipage}
	\caption{ Convergence of the derivations of the RPA@PBE optimized bond lengths from the reference values (obtained at the largest
	basis set) as a function of the basis set size. For H$_2$,  HF,  N$_2$  and CO,
	the Gaussian cc-pVXZ basis sets (X=D, T, Q, 5, abbreviated as basis set ``DZ", ``TZ", ``QZ", and ``5Z") are used. 
	For Ar$_2$, the corresponding
	aug-cc-pVXZ basis sets are used. The reference RPA@PBE bond lengths at the 5Z level are 0.7428 \AA, 0.9216 \AA, 1.1044 \AA, 1.1366 \AA\ for H$_2$, HF, N$_2$ and CO, and 3.7645 \AA\, for Ar$_2$, respectively. 
	} 
	\label{fig:basis_set_convergence_PBE_FC_cc_pVnZ_H2_HF_N2_CO_Ar2}
\end{figure}

The successful implementation of RPA analytical forces in FHI-aims allows us to relax the geometries of molecular systems
and assess the quality of RPA geometries. To begin with, we first check the convergence behavior of the RPA geometries
with respect to the basis set size. The frozen-core (FC) approximation is used for the RPA calculations below. By FC approximation,
we mean here the core electron states are excluded in the summation over occupied states in the response function calculation of Eq.~\ref{eq:chi0_realspace}, assuming that the contribution of these core states to the electron correlation energy does not change 
much across different chemical environments.
We found that the optimized RPA geometries are only marginally affected by this approximation, but one can gain a large factor in the
computation time by freezing the core electrons.
In Fig.~\ref{fig:basis_set_convergence_PBE_FC_cc_pVnZ_H2_HF_N2_CO_Ar2}, the deviations of the
optimized RPA@PBE bond lengths for five diatomic molecules (H$_2$,  HF,  N$_2$  and CO, and Ar$_2$) from their respective reference values
are plotted as a function of the basis set size .  
Here, the bond lengths calculated using cc-pV5Z (aug-cc-pV5Z for Ar$_2$) basis set are taken as the reference for each molecule. From 
Fig.~\ref{fig:basis_set_convergence_PBE_FC_cc_pVnZ_H2_HF_N2_CO_Ar2}, one can see that the optimized bond lengths do not converge monotonically
with the increase of the basis set size. However, from TZ to 5Z, the observed further changes of the bond lengths are not significant. 
At the QZ level, the optimized RPA bond lengths deviate from 
the 5Z values at most by 0.4 pm for ionically- or covalently-bonded dimers, and about 1 pm for the purely vdW-bonded dimer Ar$_2$. In the following,
we choose cc-pVQZ as the default basis set for benchmarking the quality of RPA geometries. When we find it to be necessary,
we will also check how the results change by going to larger basis sets.

 
\begin{table*} [ht!]
	\caption{\label{tab:comparison_of_results} Comparison of  RPA@PBE  bond lengths $R_e$ (in pm)  and bond angles $\varTheta_{e}$ (deg) with those
	derived from PBE and MP2, obtained using the cc-pVQZ (QZ) basis sets. The RPA geometries obtained using the cc-pV5Z (5Z) basis sets 
	are also presented. 
	 For rare gas dimers additional diffuse functions (i.e., aug-cc-pVQZ and aug-cc-pV5Z) are used in the RPA calculations.
	The reference values are taken from Ref. \onlinecite{Filip/Poul/Jeppe/stanton/2002,COOK/1975,Tang/Toennies:2003, CCCBDB} (see text
	for the actual methods), which are indicated with superscript $a$,~ $b$,~ $c$~  and $d$ respectively. The mean error (ME), 
	mean absolute error (MAE), mean percentage error (MPE) and mean absolute percentage error (MAPE) are presented for each method.
	}
\centering
	\scalebox{.9200}{
	\begin{tabular}{lcrrrrrr}
    \hline\hline 
		\multicolumn{1}{c}{\multirow{2}{*}{System}~~~~~~~} &
		\multicolumn{1}{c}{\multirow{2}{*}{~~~~~$R_e/ \varTheta_{e}$}~~~~~~} &
		\multicolumn{1}{c}{\multirow{2}{*}{~~~~~~~~~~Ref. \footnotemark[1]~~~ }} & \multicolumn{1}{c}{\multirow{2}{*}{~~~~~~~~~~RPA/QZ}~~~} & 
		\multicolumn{1}{c}{\multirow{2}{*}{~~~~~~~~~~RPA/5Z~~~}}& 
		 \multicolumn{1}{c}{\multirow{2}{*}{~~~~~~~~~~~~~PBE~~}} & 
		 \multicolumn{1}{c}{\multirow{2}{*}{~~~~~~~~~MP2\cite{CCCBDB, nwchem}~~}} 
		 & \multicolumn{1}{c}{\multirow{2}{*}{~~~~~~~~~~~~PBE0}}   \\     
		& & &  \\ 
			\hline  \\ [-0.5ex]
			\multicolumn{7}{c}{Bond lengths} \\ [3ex]
		H$_{2}$ & $R_{e}$ & 74.15~  & 0.17~  & 0.13~   &0.87~ &  $-$0.54~ &  0.30   \\

		F$_{2}$ & $R_{e}$& 141.27~ & 2.40~  &2.18~   & 0.06~ & $-$1.56~ & $-$3.71 \\
		
		N$_{2}$ & $R_{e}$& 109.76~ & 0.80~  & 0.68~  & 0.48~  & 1.28~ & $-$0.86   \\
		
		HF & $R_{e}$& 91.68~ & 0.43~  & 0.48~   & 1.25~  & 0.04~  & 0.05  \\
		
		CO & $R_{e}$& 112.82~  & 0.94~  & 0.84~   &0.71~ &	0.64~ & $-$0.59   \\
		
		CO$_{2}$ & $R_{e}$& 116.01~ & 0.93~  & 0.83~   & 1.03~  & 0.61~ & $-$0.35   \\
		
	\multirow{1}{*}{\text{HCN}}   & $R_{e}$ (H-C) &  106.53~ &0.38~  & 0.32~   & 0.93~   &0.11~  & 0.22 \\ 
		& $R_{e}$ (C-N) & 115.34~ & 0.77~ & 0.65~  & 0.40~ & 1.02~ &  $-$0.88 \\
	
	\multirow{1}{*}{\text{HNC}}   & $R_{e}$ (H-N) &  99.49~ & 0.32~  & 0.30~   & 1.00~  & 0.86~  & 0.10 \\
		& $R_{e}$ (C-N) & 116.88~ & 0.87~  & 0.77~ & 0.59~ & $-$0.05~ & $-$0.62 \\
	
	\multirow{1}{*}{\text{C$_{2}$H$_{2}$}}   & $R_{e}$ (H-C) & 106.17~ & 0.40~  & 0.32~   & 0.82~  	 & $-$0.06~ & 0.21  \\ 
		& $R_{e}$ (C-C) & 120.36~ & 0.61~ & 0.49~  & 0.28~ & 0.50~  & $-$0.79  \\
	
		\multirow{1}{*}{\text{H$_{2}$O}}   & $R_{e}$ (H-O) &  95.80~ & 0.47~   &0.46~   & 1.08~  & 0.96~  & $-$0.06 \\ 
	
		\multirow{1}{*}{\text{HNO}}   & $R_{e}$ (N-H) & 105.20~ & 0.38~ & 0.32~ & 2.90~ &  $-$0.59~  & 0.85  \\
		&$R_{e}$ (N-O)  & 120.86~ & 1.13~ & 1.04~  &$-$0.12~  & $-$0.46~  & $-$1.85 \\ 
		
	\multirow{1}{*}{\text{HOF}}   & $R_{e}$ (O-H) & 96.87~ & 	0.28~  &0.30~  & 1.07~ & $-$0.55~ & $-$0.25  \\
		& $R_{e}$ (O-F) & 143.45~ &  2.37~  & 2.21~   & 1.02~  & $-$1.52~ & $-$2.85 \\
		
	\multirow{1}{*}{\text{N$_{2}$H$_{2}$}}   & $R_{e}$ (H-N) & 102.88~ & 0.46~ & 0.41~   &  1.51~   & $-$0.18~   & 0.25 \\
		& $R_{e}$ (N-N) & 124.58~ & 1.06~  & 0.91~  &  0.17~  &0.61~ & $-$1.53 \\
	
    \multirow{1}{*}{\text{CH$_{2}$O}}   & $R_{e}$ (C-O) & 120.47~ & 0.77~  & 0.10~  &0.23~   & 0.35~ & $-$1.03 \\
		& $R_{e}$ (C-H) & 110.07~ & 0.37~  & 0.66~   &1.70~  & $-$0.15~ & 0.67	\\
		
\multirow{1}{*}{\text{C$_{2}$H$_{4}$}}   & $R_{e}$ (H-C) & 108.07~ & 0.40~  & 0.33~    & 1.02~  & $-$0.12~  & 0.29 \\ 
	    & $R_{e}$ (C-C) & 133.07~ & 0.69~   & 0.61~   & 0.15~ & $-$0.11~ & $-$0.83  \\			
		
		\multirow{1}{*}{\text{NH$_{3}$}}   & $R_{e}$ (H-N) & 101.24~ &0.47~  & 0.41~  & 0.92~  &$-$0.26~  & $-$0.05  \\ 
		
		\multirow{1}{*}{\text{CH$_{4}$}}   & $R_{e}$ (C-H) & 108.59~ &  0.50~  & 0.44~  &0.94~ &$-$0.18~  & 0.22  \\
		
		\multirow{1}{*}{\text{H$_{2}$S}}   & $R_{e}$ (H-S) & 133.56\footnotemark[2] &0.71~  &  0.56~ &1.67~ & $-$0.20~  & 0.60   	\\

      Ne$_2$ & $R_{e}$ & 310.00\footnotemark[3] & 11.60~ & 6.82~ &  $-$1.19~ & 9.20~   & 2.38   \\
	  Ar$_2$ & $R_{e}$ & 375.80\footnotemark[3] &  11.26~  & 7.86~  & 10.25~   & $-$0.40~   & 27.15  \\
	  Li$_{2}$ & $R_{e}$& 267.30\footnotemark[4] & 4.37~ & 3.11~   & 4.41~ &7.30~  &  4.76  \\
      BF & $R_{e}$&126.69\footnotemark[4]  & 0.69~ & 0.62~  &0.54~ & $-$0.25~  & $-$0.77  \\
      HCl & $R_{e}$& 127.46\footnotemark[4]  & 0.56~  & 0.47~ &1.49~  &$-$0.23~  & 0.44  \\
	  Cl$_{2}$ & $R_{e}$&  198.79\footnotemark[4] & 3.82~ & 2.76~   	& 2.49~   &$-$0.26~ & $-$0.20 \\
	  $\text{NO}^{+}$& $R_{e}$ &106.56\footnotemark[4] &0.62~  & 0.52~   & 0.36~ & 1.30~   & $-$1.25   \\
      BeH$_{2}$ & $R_{e}$&  132.64\footnotemark[4] & 0.38~   & 0.41~  & 1.03~ & 0.06~ & 0.56  \\
		
		ME & $R_{e}$& & 1.54~  & 1.16~   & 1.24~   & 0.50~ & 0.61 \\
		MAE &$R_{e}$& & 1.54~ & 1.16~   & 1.31~  & 0.96~ & 1.69  \\
		MPD & & & 0.83 \%&  0.67 \% & 0.79 \% & 0.20 \% & $-$0.01 \% \\
		MAPD &  & & 0.83 \% & 0.67 \% & 0.96 \%  &  0.58 \% & 0.85 \%  \\

			\hline  \\ [-0.5ex]
		\multicolumn{7}{c}{Bond angles} \\ [3ex]
	
	H$_2$O&  $\varTheta_{e}$  & {104.4776}$^\circ$& $-${1.0008}$^\circ$  & $-$0.7107$^\circ$    & $-${0.5571}$^\circ$   &$-${0.4648}$^\circ$  & {0.2111}$^\circ$ \\
	
		HNO & $\varTheta_{e}$ & 108.2600$^\circ$ & $-$0.3531$^\circ$   & $-$0.2839$^\circ$   & 0.3952$^\circ$ & $-$0.4711~  & 0.5308$^\circ$  \\
		
		\multirow{1}{*}{\text{HOF}}   & $\varTheta_{e}$ & 97.2000$^\circ$ & $-$0.1241$^\circ$  & $-$0.0071$^\circ$   & 0.6919$^\circ$  & 0.7962$^\circ$ & 1.8273$^\circ$  \\
		
		N$_2$H$_2$ & $\varTheta_{e}$  & {106.3000}$^\circ$ & $-${0.6066}$^\circ$   & $-$0.4516$^\circ$  &$-${0.0890}$^\circ$   &$-${0.5530}$^\circ$ & 0.5167$^\circ$  \\
		
		CH$_2$O &$\varTheta_{e}$ (HCO) & {121.9330}$^\circ$ & $-$0.0894$^\circ$  & $-$0.1907$^\circ$    &{0.1857}$^\circ$  &$-${0.1500}$^\circ$   & 0.0627$^\circ$ \\
		&$\varTheta_{e}$ (HCH) & {116.1340}$^\circ$ & 0.1793$^\circ$ & 0.3812$^\circ$    &$-${0.3714}$^\circ$   &{0.3000}$^\circ$  & $-$0.1254$^\circ$    \\
		
		C$_2$H$_4$ & $\varTheta_{e}$ (HCC)  & {121.2000}$^\circ$&{0.2341}$^\circ$  &  0.2253$^\circ$ & {0.5096}$^\circ$ &{0.1270}$^\circ$  & 0.4529$^\circ$ \\
		& $\varTheta_{e}$ (HCH)  & {117.6000}$^\circ$ &$-${0.4686}$^\circ$  & $-${0.4498}$^\circ$ & $-${1.0193}$^\circ$  & $-${0.2550}$^\circ$  & $-$0.9057$^\circ$ \\
		
		NH$_3$ & $\varTheta_{e}$  & {106.6732}$^\circ$&   $-${1.1347}$^\circ$  & $-$0.7033$^\circ$  &$-${0.8349}$^\circ$  & $-${0.2090}$^\circ$ & $-$0.0366$^\circ$  \\
		
		CH$_4$&$\varTheta_{e}$  & {109.4710}$^\circ$ & {0.0010}$^\circ$  & 0.0001$^\circ$     &{0.0000}$^\circ$ &{0.0000}$^\circ$  & $-$0.0041$^\circ$  \\

		H$_2$S & $\varTheta_{e}$  & {92.1100}$^\circ$\footnotemark[2] & 0.0455$^\circ$   & 0.0633$^\circ$   & $-${0.3828}$^\circ$   & {0.1270}$^\circ$  & 0.2117$^\circ$    \\

		ME & $\varTheta_{e}$  &  & $-$0.3016$^\circ$  &  $-$0.2114$^\circ$  & 0.1338$^\circ$  & $-$0.0684$^\circ$ & 0.2492$^\circ$ \\
		
		MAE & $\varTheta_{e}$& & {0.3852}$^\circ$ & {0.3102}$^\circ$  & {0.4579}$^\circ$ &  0.3139$^\circ$ & 0.4441$^\circ$  \\

		MPD & $\varTheta_{e}$  &  &  $-$0.29 \%  &  $-$0.18 \% & $-$0.06 \%  &  $-$0.06 \% & 0.26 \%  \\
		
		MAPD & $\varTheta_{e}$& &  0.36 \%   &   0.29 \%  &  0.36 \%  & 0.30 \% & 0.42 \% \\
		\hline\hline
		$^a$Ref.~\onlinecite{Filip/Poul/Jeppe/stanton/2002} & $^b$Ref.~\onlinecite{COOK/1975}
		& $^c$Ref.~\onlinecite{Tang/Toennies:2003} & $^d$Ref.~\onlinecite{CCCBDB} & & & 
	\end{tabular}}

\end{table*}

In Table~\ref{tab:comparison_of_results} the optimized RPA@PBE equilibrium geometries for 26 small molecules, including both bond lengths and (for poly-atomic molecules) bond angles, are presented. The PBE and MP2 geometries for the same set of molecules are also presented for comparison. 
As usual, the MP2 calculations are based on the Hartree-Fock reference states and the results are taken from Refs. \cite{nwchem,CCCBDB}. All calculations are done using the cc-pVQZ basis set (and for rare-gas dimers
the aug-cc-pVQZ basis set). For the RPA@PBE results, the cc-pV5Z (and aug-cc-pV5Z for rare-gas dimers) results are also shown,
in order to illustrate the basis set effect for RPA geometries for the entire molecular set. The reference geometries used here to benchmark
the accuracy of RPA@PBE, PBE, and MP2 are mostly taken from Refs.~\onlinecite{Filip/Poul/Jeppe/stanton/2002}. These are considered as \textit{empirical}
equilibrium geometries which are determined by a combination of experimental rotational constants and theoretical vibration-rotation interaction
constants (at the CCSD(T)/cc-pVQZ) level. The geometry values are believed to be accurate within $0.6\%$, or below 0.1 pm for bond length. 

From Table~\ref{tab:comparison_of_results}, one can see that RPA@PBE systematically  overestimates the bond lengths of small molecules, resulting in
a MAE of 1.54 pm which is essentially the same as ME, at the cc-pVQZ basis level. It appears that the basis set effect is still
appreciable at the cc-pVQZ level: When going to cc-pV5Z, the MAE (and ME) is reduced to 1.16 (1.16) pm. PBE also overestimates the bond lengths on average, 
but to a lesser extent,
with a resultant MAE comparable to RPA@PBE (at the cc-pVQZ level) but a smaller ME. In contrast, the MP2 bond lengths center around the reference values,
and do not show such a systematic overestimation behavior as observed for RPA@PBE. Furthermore, the MAE of MP2 (0.96 pm) is also noticeably smaller than
those of RPA@PBE and PBE. Close inspection reveals that the RPA@PBE bond lengths of rare-gas dimers Ne$_2$ and Ar$_2$, and alkali metal dimer Li$_2$, and 
halogen dimers F$_2$ and Cl$_2$ have particularly large errors, resulting in an overall moderate performance of RPA@PBE in determining the equilibrium
molecular structures. Such observations are consistent with the single-point energy calculations, where RPA was found to systematically underbind
molecules and solids, and adding single excitation contributions and second-order screened exchange can largely cure this deficiency 
\cite{Ren/etal:2011,Paier/etal:2012,Ren/etal:2013}. Moreover, Table~\ref{tab:comparison_of_results} suggests that 
RPA@PBE tends to overestimate the bond angles. The MAEs of the bond angles of the three methods are comparable, and the accuracy of RPA@PBE 
bond angles again is improved when increasing the basis set size from cc-pVQZ to cc-pV5Z.

We note that Burow \textit{et al.} reported in Ref.~\onlinecite{doi:10.1021/ct4008553} a much smaller bond length MAE of 0.45 pm for RPA@PBE. Careful analysis
indicates that this discrepancy mainly comes from the different molecular test sets used in the two studies. In Ref.~\onlinecite{doi:10.1021/ct4008553},
a subset of 18 molecules was used in the benchmark tests, where in our case 8 additional molecules are included. Especially the newly added molecules such as
Ne$_2$, Ar$_2$, and Cl$_2$ have particularly large errors, resulting in the final MAE three times as large for RPA@PBE. 
In our calculations, including only 18 molecules
as those in  Ref.~\onlinecite{doi:10.1021/ct4008553}, one obtains a MAE of 0.73 pm at the cc-pVQZ level, and 0.64 pm at the cc-pV5Z level. Thus our RPA@PBE error
analysis at cc-pV5Z level for a smaller test set is consistent with what is reported in Ref.~\onlinecite{doi:10.1021/ct4008553}, using the QZVPPD basis set.
In Ref.~\cite{Rekkedal/etal:2013},  Rekkedal \textit{et al.} also reported a similar MAE of 1.6 pm for RPA bond lengths. However, in the case, the larger MAE 
can be attributed to the use
of Hartree-Fock instead of PBE reference states in their calculations. In summary, we consider our calculations to be a more faithful benchmark
for RPA@PBE geometries, due to the enlarged test set and carefully checked basis set convergence behavior.

\subsection{RPA geometries and energy hierarchy for low-lying isomers of water hexamer}
Next we employ our analytical RPA force implementation to relax the structures of low-lying isomers of the gas-phase water hexamer. 
Six water molecules can arrange themselves in a number of different configurations that are very close in energy \cite{Hincapie/etal:2010}.
In particular, the four configurations with lowest energies, known as ``prism", ``book", ``cage", and ``cyclic" isomers, 
differ only by 10-20 meV/H$_2$O at most in total energy.
The correct ordering of the Born-Oppenheimer ground-state energies (i.e., without accounting for zero-point corrections and finite-temperature effect) 
among these four isomers 
has been determined by high-level quantum chemistry (MP2 \cite{Biswajit/etal:2008} and CCSD(T) \cite{Olson/etal:2007}) approaches 
as well as diffusion Monte Carlo method \cite{Biswajit/etal:2008}. These high-level calculations consistently indicate that 
the prism isomer is most stable, followed
successively by ``cage", ``book", and ``cyclic" isomers. However, most semi-local and hybrid density functional approximations
(DFAs) are unable to predict the most stable isomer, nor the correct energy ordering of the four isomers
\cite{Biswajit/etal:2008}. In Ref.~\onlinecite{Biswajit/etal:2008}, it was explained that the poor treatment of vdW interactions
is the key reason for the failure of these functionals. 
However, it was recently reported that the newly constructed SCAN meta-GGA can yield the correct energy ordering of the
four isomers \cite{Sun/etal:2016}, due to its ability to capture the mid-range vdW interactions.

In this context, it is interesting to check how the RPA performs for describing the structures and energy hierarchy of small 
water clusters. The RPA is expected to show a better performance than the conventional DFAs in describing the hydrogen-bonded
molecular networks, thanks to its ability describing vdW interactions in principle \cite{Dobson:1994,Zhu/etal:2010,Ren/etal:2011} 
and in particular the high-quality vdW coefficients it delivers \cite{Gao/Zhu/Ren:2020}. We first relax the structures of the four
water hexamers using RPA@PBE and cc-pVQZ basis set. The initial geometries for RPA@PBE relaxations were taken from
Ref.~\onlinecite{water_data_base} for the cage isomer and Ref.~\onlinecite{begdb/2008} for the three other isomers. 
The structures are relaxed until the RPA@PBE force converges within $10^{-2}$ eV/\AA.
Once the RPA geometries are obtained, we perform single-point RPA@PBE total energy calculations to determine the dissociation energies for the different isomers, 
which are defined as
 \begin{equation}
  D_{e}=-\big(E^\mathrm{RPA}(\mathrm{(H_2O)}_n)-nE^\mathrm{RPA}(\mathrm{H_{2}O)} \big)/n\, . 
 \end{equation}
In this definition,  a larger positive value of $D_{e}$ means a stronger binding of the water molecules,
and that the water cluster is more stable. Since the dissociation energies converge slower than the geometries,
here we report results obtained using several different high-quality basis sets, from which one can have a good
idea about how well the results are converged with respect to the basis set size. 
We also note that, to obtain numerically accurate RPA@PBE single point energies, we used additional 
$5g$ hydrogen-like functions (with effective charge $Z=6$) to generate extra ABFs 
(the \textit{``for\_aux"} tag in FHI-aims) \cite{Ihrig/etal:2015, Ren/etal:2021} to enhance the accuracy of LRI.

\begin{table*} [ht!]
	\caption{\label{tab:dissociation_ene_water_hexamer}  The RPA@PBE dissociation energies (in meV/H$_{2}$O) 
	of the four water hexamer structures obtained using various basis sets, for both relaxed RPA@PBE (upper panel) and 
	PBE (lower panel) geometries. The dissociation energies for the lowest-energy (prism) isomer are highlighted 
	in bold, and the energy differences between other isomers and the lowest-energy ones are indicated in
	parentheses. Both PBE and RPA@PBE geometries are relaxed using the cc-pVQZ basis set. An additional $5g (Z=6)$ ``for\_aux" basis
	function \cite{Ihrig/etal:2015} is employed to enhance the accuracy of LRI. The frozen-core approximation is used in
	all calculations.}
	\begin{tabular}{lcccl} 
		\hline\hline 
		\multicolumn{1}{c}{\multirow{2}{*}{Basis Set}} &
		\multicolumn{1}{c}{\multirow{2}{*}{Prism}} &
		\multicolumn{1}{c}{\multirow{2}{*}{Cage}} & \multicolumn{1}{c}{\multirow{2}{*}{Book}} & \multicolumn{1}{c}{\multirow{2}{*}{Cyclic}} 
		\\     
		& & & &  \\ 
		\hline 	\\ [-0.5ex]
		\multicolumn{5}{c}{RPA@PBE geometries}     \\ [3ex]
		cc-pVQZ & \bf{319.5} & 317.4(2.1) & 313.5(6.0)  & 305.3(14.2) \\
		cc-pV5Z  & \bf{300.7} & 299.2(1.5) & 297.7(3.0) &  292.5(8.2) \\
		NAO-VCC-4Z & \bf{300.6} & 299.5(1.1) & 298.1(2.5) & 292.2(8.4) \\
		aug-cc-pVQZ & \bf{306.6} & 304.7(1.9)  & 302.4(4.2) & 295.9(10.7) \\
		aug-cc-pV5Z & \bf{301.0} & 299.3(1.7) & 297.1(3.9) & 290.9(10.1)   \\
		CBS(aQZ,a5Z) & \bf{295.1} & 293.6(1.5) & 291.5(3.6) & 285.7(9.5)   \\
		[2ex]
		\cline{2-5} \\[1ex]
		\multicolumn{5}{c}{PBE geometries}  \\ [3ex]
	
 		cc-pVQZ & \bf{311.2} & 307.8(3.4) & 302.5(8.7)   &  293.5(17.7)    \\
 		cc-pV5Z  & \bf{291.5} & 288.3(3.2)  & 285.0(6.5) &  278.7(12.8)  \\
 		NAO-VCC-4Z &  \bf{291.8} & 289.0(2.8) & 285.9(5.9) & 279.1(12.7) \\
 		aug-cc-pVQZ & \bf{297.9} & 294.4(3.5)  & 290.2(7.7)  & 282.8(15.1) \\
 	    aug-cc-pV5Z & \bf{291.7} & 288.2(3.5) & 284.1(7.6) & 276.5(15.2)   \\
 	    CBS(aQ,a5Z) & \bf{285.2} & 281.7(3.5) & 277.7(7.5) & 269.9(15.3)   \\
		\hline\hline
	\end{tabular}
\end{table*} 

In Table~\ref{tab:dissociation_ene_water_hexamer} the RPA@PBE dissociation energies for the four water hexamers 
obtained using Gaussian cc-pVQZ, cc-pV5Z, aug-cc-pVQZ (aQZ), aug-cc-pV5Z (a5Z), and NAO-VCC-4Z \cite{IgorZhang/etal:2013} basis sets are presented.
Also presented are the dissociation energies at the complete basis set (CBS) limit, obtained via the two-point extrapolation procedure \cite{Helgaker/etal:1997} using aQZ/a5Z data points. 
Under this procedure, $D_{e}$ at the CBS limit is given by,
\begin{equation}
    D_{e}(\text{CBS}) = \frac{4^3 D_e(\text{aQZ})-5^3 D_e(\text{a5Z})}{4^3-5^3} \, .
\end{equation}
The basis set superposition error (BSSE) is not corrected in all these calculations, as the BSSE should get diminishingly small as one goes to
the CBS limit.
For comparison, in addition to the results obtained using the RPA@PBE geometries, we also present the RPA@PBE dissociation energies for
the same set of isomers and basis functions using the PBE geometries. 
First of all, one can see from Table~\ref{tab:dissociation_ene_water_hexamer} that, for all basis sets, the correct energy ordering of the
four isomers is obtained. At the level of 5Z/a5Z, the calculated RPA@PBE dissociation energies differ from their counterparts at the CBS
limit by 6-7 meV per H$_2$O molecule and the energy differences between the isomers differ within 1 meV. Interestingly, results of comparable quality are obtained
using the NAO-VCC-4Z basis set \cite{IgorZhang/etal:2013}. These are numerical orbitals but generated according to a similar ``correlation consistent"
principle as the Dunning's basis sets \cite{Dunning:1989}. Here, it seems the $4Z$-level NAO basis set can yield RPA results of comparable quality
as the $5Z$ Gaussian basis sets.

As mentioned above, in Table~\ref{tab:dissociation_ene_water_hexamer}, 
we also investigate the effect of a different geometry (here, the geometry obtained by PBE) on the RPA energies for water hexamer. 
Compared to the results obtained using the PBE geometries, the RPA dissociation energies obtained using the RPA geometries 
are bigger in magnitude, meaning that
the water clusters settled down in the RPA geometries have lower energies. This is exactly what one would expect, since the RPA geometries
are local minima of the RPA potential energy surface, while the PBE geometries are not. Comparing the two sets of results, the water hexamers in their RPA
geometries are lower in energy by about 10 meV/H$_2$O than their counterparts in PBE geometries. The energy differences between 
the different isomers are noticeably larger in PBE geometries, although the energy ordering among the four isomers does not change. 
We expect that such structural effects, although not causing a qualitative difference in water hexamers, 
might have implications for bigger water clusters and bulk water \cite{Yao/Kanai:2021}.

In Table~\ref{tab:dissociation_ene_water_hexamer_comparison_wd_Santra_paper}, we compare the RPA@PBE dissociation energies for
the four isomers of water hexamer with those obtained by quantum chemistry approaches and conventional DFAs. The RPA@PBE results
are obtained in their own geometries and extrapolated to the CBS limit. One can see that the energy
differences between different isomers yielded by RPA@PBE are in rather good agreement with the CCSD(T) results. In particular,
RPA@PBE seems to show a better performance than MP2, which tends to underestimate the energy differences of the cage, book, and cyclic
isomers from the prism isomer. However, RPA@PBE shows a substantial underestimation of the dissociation energies themselves -- as large as
50 meV per water molecule. Furthermore, the energy separations between different isomers remain to be underestimated within RPA@PBE,
despite its improvement over MP2. The fact that RPA underbinds molecules is well known, and in the past it has been demonstrated
that a correction term arising from single excitations helps alleviate this problem \cite{Ren/etal:2011,Paier/etal:2012,Ren/etal:2013}.
In Table~\ref{tab:dissociation_ene_water_hexamer_comparison_wd_Santra_paper} we also present the results obtained by RPA plus renormalized
single excitations (rSE) correction. It can be seen that RPA+rSE results are in excellent agreement with the CCSD(T) results, both
for the dissociation energies and for the energy differences between different isomers. 

As post-KS methods, the results of RPA and RPA+rSE necessarily depend on the starting point, i.e., the orbitals and orbital energies generated
from a preceding self-consistent field (SCF) calculation. To check this dependence, we performed single-point RPA and RPA+rSE energy calculations
based on the hybrid PBE0 \cite{Perdew/Ernzerhof/Burke:1996,Ernzerhof/Scuseria:1999} functional using the optimized RPA@PBE geometries.
As can be seen from the results presented also in Table~\ref{tab:dissociation_ene_water_hexamer_comparison_wd_Santra_paper}, the dissociation energies
of RPA@PBE0 for the water clusters get larger (correctly) and improve over RPA@PBE. On the contrary, the (RPA+rSE)@PBE0 dissociation energies 
get smaller compared to their (RPA+rSE)@PBE counterparts, and show a larger derivation from the CCSD(T) results. Nevertheless, the energy differences yielded (RPA+rSE)@PBE0 are still
in rather good agreement with the CCSD(T) data. This behavior isn't surprising, as the magnitude of the rSE correction
depends on the preceding SCF reference, owing to the fact that the rSE contribution originates from the difference between exact-exchange and KS exchange potentials \cite{Ren/etal:2011,Ren/etal:2013}.
 With hybrid functional starting points,
the rSE correction will necessarily get smaller than that with pure local/semilocal functional starting points.

In essence, these benchmark results indicate that RPA+rSE might
be a promising affordable approach to study properties of water clusters. In contrast, with the exception of SCAN meta-GGA functional
\cite{Sun/etal:2016}, conventional DFAs, even with vdW corrections, can barely describe the energy hierarchy among different water hexamers,
as also indicated in Table~\ref{tab:dissociation_ene_water_hexamer_comparison_wd_Santra_paper}.

   \begin{table*} [ht]
    	\caption{ \label{tab:dissociation_ene_water_hexamer_comparison_wd_Santra_paper}
    	Comparison of dissociation energies of the four water hexamers (in meV/H$_2$O) given by RPA@PBE and various other electronic structure approaches. The RPA@PBE, RPA@PBE0, (RPA+rSE)@PBE  and (RPA+rSE)@PBE0 results at the CBS(aQZ,a5Z) limit are obtained in this work. 
    	The CCSD(T) results are taken from Ref.~\cite{Olson/etal:2007} and the results of all other approaches
    	are taken from Ref.~\cite{Biswajit/etal:2008}. Here the RPA and RPA+rSE energetic calculations 
    	are done using RPA@PBE geometries, the CCSD(T) calculations are done using MP2 geometries, and 
    	all other types of calculations are done using the geometries generated by the same method. 
    	Similar to Table~\ref{tab:dissociation_ene_water_hexamer}, the dissociation energies for 
    	the lowest-energy isomer are highlighted in bold, and the energy difference between other isomers and 
    	the lowest-energy one are indicated in parentheses. }
    	\begin{tabular}{lllll} 
    		\hline\hline 
    	\multicolumn{1}{c}{\multirow{2}{*}{~~~~~Method~~~~~~~~~~~~}} &
    		\multicolumn{1}{c}{\multirow{2}{*}{Prism}} &
    		\multicolumn{1}{c}{\multirow{2}{*}{Cage}} & \multicolumn{1}{c}{\multirow{2}{*}{Book}} & \multicolumn{1}{c}{\multirow{2}{*}{Cyclic}} 
    		 \\\\
    		\hline 	
    		
    	RPA@PBE & \bf{295.1} & 293.6(1.5) & 291.5(3.6) & 285.7(9.5) \\
    	RPA@PBE0 & \bf{308.5} & 306.9(1.6) & 305.5(3.0) & 300.5(8.0)   \\
    	(RPA+rSE)@PBE & \bf{344.5} & 342.3(2.2) & 335.7(8.8)  & 325.4(19.1) \\
    	(RPA+rSE)@PBE0 & \bf{334.6} & 332.6(2.0) & 329.2(5.4) & 321.1(13.5)   \\
    		MP2$^b$ & \textbf{332.3} & 331.9(0.4) &330.2(2.1) & 324.1(8.2) \\
    		CCSD(T)$^a$ & \textbf{347.6} & 345.5(2.1) & 338.9(8.7) & 332.5(15.1) \\ 
    		PBE$^b$ & 336.1(9.5) & 339.4(6.2) & \textbf{345.6} & 344.1(1.5)\\
    		PBE0$^b$ &  322.9(8.0) & 325.3(5.7) & \textbf{330.9} & 330.8(0.1)
    		\\
    		BLYP$^b$& 273.6(16.2) & 277.4(12.4)& 287.5(2.3)& \bf{289.8}\\
    		PBE+vdW$^b$ & 377.8(2.3)& \bf{380.1}& 377.8(2.3.) &  367.3(12.8)\\
    		PBE0+vdW$^b$ & 360.6(1.3)& \bf{361.9} &359.2(2.7) & 351.4(10.5) \\
    		BLYP+vdW$^b$& \bf{359.9} & 359.7(0.2)&356.3(3.6) & 344.8(15.1)  \\
    		\hline\hline  
    		$^a$Ref~\onlinecite{Olson/etal:2007} & & $^b$Ref~\onlinecite{Biswajit/etal:2008} & &\\
    	\end{tabular} 
    \end{table*} 

   \section{Conclusion} 
    In conclusion, we have derived and implemented a formalism for analytical RPA force calculations based on LRI and DFPT. Our
    current implementation in FHI-aims features a canonical $O(N^5)$ scaling but a lower scaling is possible by exploiting the
    sparsity offered by LRI. Benchmark calculations against forces obtained by the finite-difference method and RPA geometries reported
    in the literature show that our RPA force implementation is highly accurate. By studying the geometries of a representative set of 26 molecules
    with various bonding characteristics, we found that the usual post-KS RPA approach shows a tendency to systematically overestimate the bond lengths.
    We then looked at the energy hierarchy of the low-lying isomers of water hexamer, and found that RPA@PBE is able to yield correct energy ordering
    for the lowest four isomers in both PBE and RPA geometries. However, the energy separations between different isomers are appreciably smaller
    in their RPA geometries. Finally, by adding the rSE corrections to RPA, both the dissociation energies of the water clusters and their energy differences
    get substantially improved. 
    \\
    
    \section*{Acknowledgments}
     The work is supported by National Natural Science Foundation of China (Grant Nos. 11874036, 11874335),  Local Innovative and Research Teams Project of Guangdong 
     Pearl River Talents Program (2017BT01N111), Shenzhen Basic Research Project (JCYJ20200109142816479), 
     and Max Planck Partner Group for
    \textit{Advanced Electronic Structure Methods}. 

\newpage

	

\bibliography{./CommonBib}

\end{document}